\documentclass[prb,showpacs,twocolumn]{revtex4}%
\usepackage{amsmath}
\usepackage{amssymb}
\usepackage{graphicx}
\usepackage{bm}
\usepackage{amsfonts}%
\providecommand{\U}[1]{\protect\rule{.1in}{.1in}}

\newcommand{\be}{\begin{eqnarray}}
\newcommand{\en}{\end{eqnarray}}
\newcommand{\ben}{\begin{eqnarray}}
\newcommand{\enn}{\end{eqnarray}}
\newcommand{\beq}{\begin{eqnarray}}
\newcommand{\eeq}{\end{eqnarray}}

\newcommand{\benon}{\begin{eqnarray*}}
\newcommand{\ennon}{\end{eqnarray*}}
\begin{document}
\preprint{APS/123-QED}
\title{Theory of spin current in chiral helimagnet}
\author{I.~G.~Bostrem,$^{1}$ Jun-ichiro Kishine,$^{2}$ and A.~S.~Ovchinnikov$^{1}$ }
\affiliation{$^{1}$Department of Physics, Ural State University, Ekaterinburg, 620083 Russia,}
\affiliation{$^{2}$Department of Basic Sciences, Kyushu Institute of Technology, Kitakyushu
804-8550, Japan}
\date{\today}

\begin{abstract}
We give detailed description of the transport spin current in the chiral
helimagnet. Under the static magnetic field applied perpendicular to the
helical axis, the magnetic kink crystal (chiral soliton lattice) is formed.
Once the kink crystal begins to move under the Galilean boost, the
spin-density accumulation occurs inside each kink and there emerges periodic
arrays of the induced magnetic dipoles carrying the transport spin current.
The coherent motion of the kink crystal dynamically generates the spontaneous
demagnetization field. This mechanism is analogous to the
D\"{o}ring-Becker-Kittel mechanism of the domain wall motion in ferromagnets.
To describe the kink crystal motion, we took account of not only the
tangential $\varphi$-fluctuations but the longitudinal $\theta$-fluctuations
around the helimagnetic configuration. Based on the collective coordinate
method and the Dirac's canonical formulation for the singular Lagrangian
system, we derived the closed formulae for the mass, spin current and induced
magnetic dipole moment accompanied with the kink crystal motion. To
materialize the theoretical model presented here, symmetry-adapted material
synthesis would be required, where the interplay of crystallographic and
magnetic chirality plays a key role there.

\end{abstract}

\pacs{Valid PACS appear here}
\maketitle

\address{
$^1$Department of Physics, Ural State University, Ekaterinburg, 620083 Russia\\
$^2$Department of Fundamental Sciences, Kyushu Institute of Technology, Kitakyushu 804-8550, Japan
}

\section{INTRODUCTION}

The core problem in the multidisciplinary field of spintronics is how to
create, transport, and manipulate spin currents.\cite{Zutic04} The key notions
include the current-driven spin-transfer
torque\cite{Slonczewski96,Berger96,Stiles02,Stiles02b,Tatara-Kohno04} and
resultant force acting on a domain wall (DW)\cite{Aharonov-Stern,Bazaliy} in
metallic ferromagnetic/nonmagnetic multilayers, the dissipationless spin
currents in paramagnetic spin-orbit coupled
systems,\cite{Rashba60,Murakami03,Sinova04} and magnon transport in textured
magnetic structures.\cite{Bruno05} A fundamental query behind the issue is how
to describe transport spin currents.\cite{Rashba05} To make clear the meaning
of the spin currents, we need to note the spin can appear in the macroscopic
Maxwell equations only in the form of spin magnetization. In this viewpoint,
the spin current is understood as the deviation of the spin projection from
its equilibrium value. An emergence of the coherent collective transport in
\textit{non-equilibrium} state is then a manifestation of the dynamical
off-diagonal long range order (DODLRO).\cite{Fomin91,Volovik07}

On the other hand, the physical currents are classified into two categories,
i.e., the gauge current originating from the gauge invariance and the inertial
current originating from the Galilean invariance. The electric current is the
gauge current, where the electric charge is coupled to the electromagnetic
U(1) gauge field. The electromagnetic field is a physical gauge field that has
its own dynamics, i.e., we know the electromagnetic field energy. Then, the
charge current $j_{i}$ and the charge density $\rho$ are related via the
continuity equation $\partial\rho/\partial t=-\partial j_{i}/\partial x_{i}$.
On the other hand, a typical example of the inertial current is the momentum
current in a classical ideal fluid, where the momentum current $\Pi_{ij}$
satisfies the continuity equation, $\partial\left(  \rho v_{i}\right)
/\partial t=-\partial\Pi_{ij}/\partial x_{j},$ and given by $\Pi_{ij}%
=P\delta_{ij}+\rho v_{i}v_{j}$ with $P$ being equilibrium
pressure.\cite{LLFluid} The non-equilibrium current is described by $\rho
v_{i}v_{j}$. In the spin current problem, at present, we have no known gauge
field directly coupled to the spin current. Therefore, a promising candidate
is the \textit{inertial current of the magnetization}.

Historically, D\"{o}ring\cite{Doring48} pointed out that the longitudinal
component of the slanted magnetic moment inside the Bloch DW emerges as a
consequence of translational motion of the DW. An additional magnetic energy
associated with the resultant demagnetization field is interpreted as the
kinetic energy of the wall. This idea was simplified by Becker\cite{Becker50}
and Kittel.\cite{Kittel50} Recent progress of material synthesis sheds new
light on this problem. In a series of magnets belonging to chiral space group
without any rotoinversion symmetry elements, the crystallographic chirality
gives rise to the asymmetric Dzyaloshinskii interaction that stabilizes either
left-handed or right-handed chiral magnetic structures.\cite{Dzyaloshinskii58}
In these chiral helimagnets, magnetic field applied perpendicular to the
helical axis stabilizes a periodic array of DWs with definite spin chirality
forming kink crystal or chiral soliton
lattice.\cite{Kishine_Inoue_Yoshida2005}

We recently proposed a new way to generate a spin current in the chiral
helimagnets with magnetic field applied in the plain of rotation of
magnetization.\cite{BKO08} The mechanism is quite analogous to the
D\"{o}ring-Becker-Kittel mechanism. We showed that the periodic spin
accumulation occurs as a dynamical effect caused by the moving magnetic kink
crystal (chiral soliton lattice) formed in the chiral helimagnet under the
static magnetic field applied perpendicular to the helical axis. The current
is inertial flow triggered by the Galilean boost of the kink crystal. An
emergence of the transport magnetic currents is then a consequence of the
dynamical off-diagonal long range order along the helical axis.

{In this paper, we give an extension of the results touched on in Ref.
\cite{BKO08}. In Sec. II, we give an overview of basic properties of the
chiral magnets that materialize the theoretical model considered in this
paper. In Sec. III, we present standard description of the kink crystal
formation, and the vibrational modes around the kink crystal state. In Sec.
IV, we apply the collective coordinate method to the moving kink crystal that
makes clear the physical meaning of the mass and the magnon current carried by
the moving system. In Sec. V, we perform quantitative estimates of the mass,
magnetic current, and net magnetization induced by the movement. In Sec. VI,
we discuss issues closely related to the present problem, including the
background spin current problem, spin supercurrent in the superfluid $^{3}$He,
and experimental aspects of our effects. Finally, we summarize the paper in
Sec. VII. }

\section{Chiral helimagnet}

In this section, we briefly review basic properties of chiral helimagnets that
materialize our theoretical model. Recent progress of material synthesis
promotes systematic researches on a series of magnets belonging to chiral
space group without any rotoinversion symmetry
elements.\cite{Kishine_Inoue_Yoshida2005} In the chiral magnets, the
crystallographic chirality possibly gives rise to the asymmetric
Dzyaloshinskii interaction that stabilizes the chiral helimagnetic structure,
where either left-handed or right-handed magnetic chiral helix is
formed.\cite{Dzyaloshinskii58} As we will see, in the chiral helimagnets,
magnetic field applied perpendicular to the helical axis stabilizes a periodic
array of DWs with definite spin chirality forming kink crystal or chiral
soliton lattice.\cite{Kishine_Inoue_Yoshida2005}

The chiral helimagnetic structure is an incommensurate magnetic structure with
a single propagation vector $\mathbf{k}_{0}=(0,0,k)$. The space group
$\mathcal{G}$ consists of the elements $\{g_{i}\}$. Among them, some elements
leave the propagation vector $\mathbf{k}_{0}=(0,0,k)$ invariant, i.e., these
elements form the little group $G_{\mathbf{k}_{0}}$.\cite{Izumov,Kovalev} The
magnetic representation\cite{Kovalev} $\Gamma_{mag}$ is written as
$\Gamma_{mag}=\Gamma_{\mathrm{perm}}\otimes\Gamma_{\mathrm{axial}}$, where
$\Gamma_{\mathrm{perm}}$ and $\Gamma_{\mathrm{axial}}$ represent the Wyckoff
permutation representation and the axial vector representation, respectively.
Then, $\Gamma_{mag}$ is decomposed into the non-zero irreducible
representations of $G_{\mathbf{k}_{0}}$. The incommensurate magnetic structure
is determined by a \textquotedblleft magnetic basis frame\textquotedblright of
an axial vector space and the propagation vector $\mathbf{k}$ . In specific
magnetic ion, the decomposition becomes $\Gamma_{mag}=\sum_{i}n_{i}\Gamma_{i}%
$, where $\Gamma_{i}$ is the irreducible representations of $G_{\mathbf{k}%
_{0}}$. Then, we have two cases leading to the chiral helimagnetic magnetic
structure. Case I: The magnetic moments are described by two independent
one-dimensional representations that form two-dimensional basis frames, or
Case II: The magnetic moments are described by a single two-dimensional
representations that form two-dimensional basis frames. In these cases, the
symmetry condition allows the chiral helimagnetic structure to be realized.
Then, the structure is stabilized by the generalized Dzyaloshinskii
interaction. The generalized Dzyaloshinskii interaction means symmetry-adapted
anti-symmetric exchange interaction, not restricted to conventional
Dzyaloshinskii-Moriya (DM) interaction caused by the on-site spin-orbit
coupling and the inter-site exchange interactions. The presence of this term
is justified by the existence of the Lifshitz invariant\cite{Dzyaloshinskii64}
for the little group $G_{\mathbf{k}_{0}}$.

Among the inorganic chiral helimagnets, the best known example is the metallic
helimagnet MnSi (${T}_{\text{c}}\simeq30$K) that belongs to the cubic space
group $P$2$_{1}$3($a=4.558$\AA ).\cite{Ishikawa76} The metallic helimagnet
Cr$_{1/3}$NbS$_{2}$ (${T}_{\text{c}}\simeq120$K)\ belongs to the hexagonal
space group $P$6$_{3}$22 ($a=5.75$\AA $,$ $c=12.12$%
\AA ).\cite{Moriya-Miyadai82} The insulating copper metaborate, CuB$_{2}%
$O$_{4}$ (${T}_{\text{c}}\simeq10$K)\ has a larger unit cell and belongs to
the tetragonal space group $I\bar{4}2d$ ($a=11.48$\AA , $c=5.620$%
\AA ).\cite{Roessli01,Kousaka07} As examples of molecular-based magnets, the
structurally characterized green needle, [Cr(CN)$_{6}$][Mn{($S$ or $R$%
)-pnH}(H$_{2}$O)]H$_{2}$O (${T}_{\text{c}}\simeq38$K),\ belongs to the
orthorhombic space group $P2_{1}2_{1}2_{1}$ ($a=7.628$\AA , $b=14.51$\AA ,
$c=14.93$\AA ). The yellow needle, K$_{0.4}$[Cr(CN)$_{6}$][Mn($S$%
)-pn]($S$)-pnH$_{0.6}$: (($S$)-pn = ($S$)-1,2-diaminopropane) (${T}_{\text{c}%
}\simeq53$K), belongs to the hexagonal space group $P6_{1}$ ($a=14.77$\AA ,
$c=17.57$\AA ).\cite{Kishine_Inoue_Yoshida2005} From the symmetry-based
viewpoints, these space groups are all eligible to realize\ the chiral
helimagnetic order.

\section{Kink crystal and vibrational modes around the kink-crystal state}

\begin{figure}
[h]
\begin{center}
\includegraphics[
height=1.1649in,
width=2.0877in
]%
{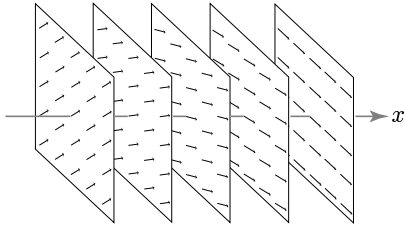}%
\caption{Schematic view of the model chiral helimagnet considered here. }%
\label{fig:chiral_helimagnet}%
\end{center}
\end{figure}

As shown in Fig.~\ref{fig:chiral_helimagnet}, we consider a system of the
chiral helimagnetic chains described by the model Hamiltonian
\begin{align}
\mathcal{H}  &  =-J\sum_{<i,j>}\mathbf{S}_{i}\cdot\mathbf{S}_{j}\nonumber\\
&  +\mathbf{D}\cdot\sum_{<i,j>}\mathbf{S}_{i}\times\mathbf{S}_{j}%
-\mathbf{\tilde{H}}\cdot\sum_{i}\mathbf{S}_{i}, \label{lattH}%
\end{align}
where the first term represents the ferromagnetic coupling with the strength
$J>0$ between the nearest neighbor spins $\mathbf{S}_{i}$ and $\mathbf{S}_{j}%
$. The second term represents the parity-violating Dzyaloshinskii interaction
between the nearest neighbors, characterized by the the mono-axial vector
$\mathbf{D}=D\hat{\mathbf{e}}_{x}$ along a certain crystallographic chiral
axis (taken as the $x$-axis). The third term is the Zeeman coupling with the
magnetic field $\mathbf{\tilde{H}}=g\mu_{B}H\hat{\mathbf{e}}_{y}$ applied
\textit{perpendicular} to the chiral axis. {When we treat the model
Hamiltonian (\ref{lattH}), we implicitly assume that the magnetic atoms form a
cubic lattice and the uniform ferromagnetic coupling exists between the
adjacent chains to stabilize the long-range order. Then, the Hamiltonian
(\ref{lattH}) is interpreted as a quasi one-dimensional model based on the
interchain mean field picture.\cite{SIP} }

When $H=0$, the long-period incommensurate helimagnetic structure is
stabilized with the definite chirality (left-handed or right-handed) fixed by
the direction of the mono-axial $\mathbf{D}$-vector. The Hamiltonian
(\ref{lattH}) is the same as the model treated by Liu\cite{Liu73} except that
we ignore the single-ion anisotropy energy. Once we take into account the
easy-axis type anisotropy term, $-K\sum_{i}(S_{i}^{x})^{2}$, the mean field
ground state configuration becomes either the chiral helimagnet for
$K<D^{2}/J$, or the Ising ferromagnet for $K>D^{2}/J$. In this paper, we
assume $K=0$ and left an effect of $K$ for a future study.%

\begin{figure}
[h]
\begin{center}
\includegraphics[
height=1.9917in,
width=3in
]%
{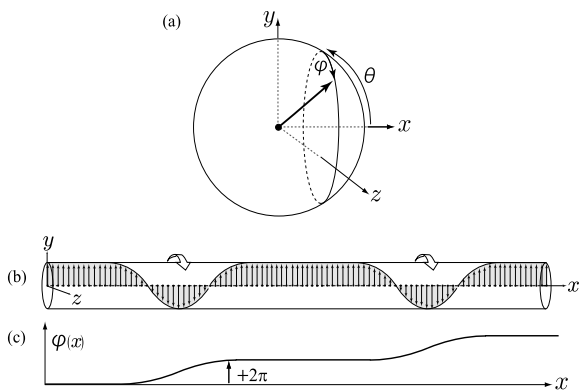}%
\caption{(a) Polar coordinates in the laboratory frame. (b) Formation of the
magnetic kink crystal in the chiral helimagnets under the transverse magnetic
field, and (c) concomitant phase modulation. In (b), we depict a linear array
of the spins along one chiral axis that is ferromagnetically coupled to the
neighboring arrays. }%
\label{CSL}%
\end{center}
\end{figure}

Taking the semiclassical parametrization of Heisenberg spins in the continuum
limit $\mathbf{S}\left(  x\right)  =S(\cos\theta\left(  x\right)  ,\sin
\theta\left(  x\right)  \cos\varphi\left(  x\right)  ,\sin\theta\left(
x\right)  \sin\varphi\left(  x\right)  )$ by using the slowly varying polar
angles $\theta(x)$ and $\varphi(x)$ [see Fig.~\ref{CSL}(a)], the Hamiltonian
acquires the form
\begin{align}
&  \mathcal{H}\left[  {\varphi}\left(  x\right)  ,{\theta}\left(  x\right)
\right]  \nonumber\\
\;\; &  =JS^{2}\int_{0}^{L}dx\left[  {\frac{1}{2}}\left\{  {\partial}%
_{x}{\theta}\left(  x\right)  \right\}  ^{2}+{\frac{1}{2}}\sin^{2}%
\theta\left\{  {\partial}_{x}{\varphi}\left(  x\right)  \right\}  ^{2}\right.
\nonumber\\
&  \;\;\left.  -q_{0}\sin^{2}\theta\left(  x\right)  {\partial}_{x}{\varphi
}\left(  x\right)  -m^{2}\sin\theta\left(  x\right)  \cos\varphi\left(
x\right)  \right]  ,\label{cont1}%
\end{align}
where $m=\sqrt{{\tilde{H}}/{{{J}S}}}$, and $L$ denotes the linear dimension of
the system. From now on, all distances are measured in the lattice unit
$a_{0}$. The helical pitch in the zero field ($m=0$) is given by $q_{0}=D/J$.%

The magnetic kink crystal phase is described by the stationary soliton
solution minimizing $\mathcal{H}$, $\theta=\pi/2$ and $\cos\left[
{{\varphi_{0}(x)}}/{2}\right]  =\mathrm{sn}\left[  {m}x/{\kappa}%
,\kappa\right]  $, where $\mathrm{sn}$ is the Jacobi elliptic function with
the elliptic modulus $\kappa$ ($0<\kappa^{2}<1$%
).\cite{Dzyaloshinskii64,Rubinstein70} This solution corresponds to a periodic
regular array of the magnetic kinks with the "topological charge" density
$\partial_{x}{\varphi_{0}\left(  x\right)  =2}\dfrac{m}{\kappa}\mathrm{dn}%
\left(  {\dfrac{m}{\kappa}}x,\kappa\right) $ as shown in Figs.~\ref{CSL}(b)
and (c). The elliptic modulus $\kappa$ is found from the minimization of
energy per unit length that yields $\kappa/m={4E(\kappa)/\pi q_{0}}%
$.\cite{Dzyaloshinskii64} The period of the soliton lattice is given by
\begin{equation}
\ell_{\text{kink}}=\frac{{2\kappa K(\kappa)}}{{m}}=\frac{{8K(\kappa)}%
E(\kappa)}{\pi q_{0}}, \label{period}%
\end{equation}
where $K(\kappa)$ and $E(\kappa)$ denote the elliptic integrals of the first
and second kind, respectively. The period increases from $2\pi/q_{0}$ to
infinity as ${\kappa}$ increases from zero to unity. In the limit of
${\kappa\rightarrow0}$, the $\mathrm{sn\,}${function approaches }${\sin}$ and
$\kappa/m\mathrm{\,}{\rightarrow2/}q_{0}$, i.e. ${\varphi_{0}(x)=q}_{0}x$ as
it should be in the case of zero field.

In the Hamiltonian (\ref{cont1}), the exchange processes favor the
incommensurate (IC) chiral helimagnetic order, while the Zeeman term favors
the commensurate (C) phase. The C-IC transition occurs at $\kappa=1$ provided
$E(1)=1$, and the critical value of ${m}$ is given by $\pi q_{0}/4m_{c}=1$.
\cite{Bak,Bulaevskii,Pokrovskii} The critical field strength $\tilde{H}_{c}$
is determined from $\sqrt{{\tilde{H}}/{\tilde{H}_{c}}}={{\kappa}/{E(\kappa)}}$.

Next, we consider the fluctuations around the kink crystal state. The studies
of collective excitations in the system have been focused on the phasons
($\phi$-mode) presenting bending waves of domain walls of the soliton
lattice.\cite{McMillan} In our analysis we are interested in the $\theta
$-modes also. We derive the spectrum of elementary exciations holding the
scheme outlined in Ref.\cite{Izyumov-Laptev86}

In what follows, it is convenient to work with the dimensionless coordinate%
\begin{equation}
\bar{x}=\dfrac{m}{\kappa}x=2K\left(  \kappa\right)  \dfrac{x}{\ell
_{\text{kink}}}=\frac{\pi}{4E\left(  \kappa\right)  }q_{0}x.
\end{equation}
We introduce $\bar{L}=mL/\kappa$ and $\bar{q}_{0}=\kappa q_{0}/m,$ and rewrite
the Hamiltonian (\ref{cont1}) as%
\begin{equation}
\mathcal{H}=JS^{2}\,\frac{m}{\kappa}\,\overline{\mathcal{H}}=JS^{2}%
\,\frac{2K\left(  \kappa\right)  }{\ell_{\text{kink}}}\,\overline{\mathcal{H}%
},
\end{equation}
where the dimensionless Hamiltonian $\overline{\mathcal{H}}$ is defined by
\begin{widetext}
\begin{eqnarray}
\overline{\mathcal{H}} &=&\int_{0}^{\bar{L}}d{\bar{x}}\left[ {\frac{1}{2}}%
\left\{ {\partial _{\bar{x}}\theta }\left( {\bar{x}}\right) \right\} ^{2}+{%
\frac{1}{2}}\sin ^{2}\theta \left( {\bar{x}}\right) \left\{ {\partial _{{x}%
}\varphi }\left( {\bar{x}}\right) \right\} ^{2}
-\bar{q}_{0}\sin ^{2}\theta \left( {x}\right) {\partial }_{{_{\bar{x%
}}}}{\varphi }\left( {\bar{x}}\right) -\kappa ^{2}\sin \theta \left( {\bar{x}%
}\right) \cos \varphi \left( {\bar{x}}\right) \right] .
\label{cont3}
\end{eqnarray}%
\end{widetext}

\begin{figure}
[h]
\begin{center}
\includegraphics[
height=3.186in,
width=2.3644in
]%
{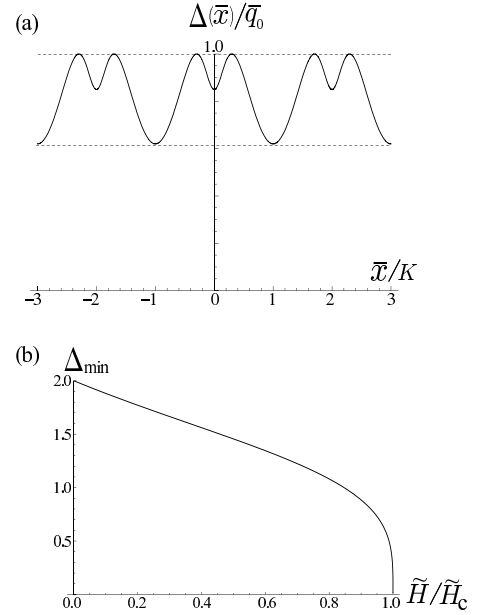}%
\caption{(a) Spatial variatoin of the gap function $\Delta\left(  \bar
{x}\right)  $ for the $\vartheta$-mode. (b) The minimum gap $\Delta_{\min}$ is
shown as a function of $\tilde{H}/\tilde{H}_{c}$.}%
\label{gapfunction}%
\end{center}
\end{figure}

As the magnetic field increases from $H=0$ to $H=H_{\text{c}}$, the parameter
${\bar{q}_{0}}=4E(\kappa)/\pi$ monotonously decreases from ${\bar{q}_{0}=}2$
to ${\bar{q}_{0}=4/\pi}${$\simeq$}${1.273}$. The fluctuations consist of the
vibrational (phonon) modes and the translational mode, that are separately
treated. In this section, we examine the phonon modes. We write
\begin{equation}
\varphi(\bar{x})=\varphi_{0}(\bar{x})+v(\bar{x}),\;\;\;\;\;\theta(\bar
{x})=\dfrac{\pi}{2}+u(\bar{x})
\end{equation}
and expand (\ref{cont3}) up to $u^{2}$ and $v^{2}.$ Then we have
$\overline{\mathcal{H}}$\ $=\int_{0}^{\bar{L}}d\bar{x}(\overline{\mathcal{H}%
}_{0}+\overline{\mathcal{H}}_{u}+\overline{\mathcal{H}}_{v}+\overline
{\mathcal{H}}_{\mathrm{int}})+{\mathcal{O}}\left(  u^{2},v^{2}\right)  ,$
where ${\mathcal{H}}_{0}$ corresponds to the stationary solution. The
interaction part contains $-u^{2}(\partial_{\bar{x}}v)^{2}/2$ and $u^{4}%
$\ terms that are neglected here. The vibrational term ${\mathcal{V}}=\int
_{0}^{\bar{L}}d{x}\left(  \overline{\mathcal{H}}_{u}+\overline{\mathcal{H}%
}_{v}\right)  $\ is given by $\overline{\mathcal{H}}_{u}=u\widehat
{{\mathcal{L}}}_{u}u$, and $\overline{\mathcal{H}}_{v}=v\widehat{{\mathcal{L}%
}}_{v}v,$ where the differential operators, $\widehat{{\mathcal{L}}}_{v}$ and
$\widehat{{\mathcal{L}}}_{u},$ are defined by
\begin{equation}
\left\{
\begin{array}
[c]{l}%
\widehat{{\mathcal{L}}}_{v}=-{\dfrac{1}{2}}\partial_{\bar{x}}^{2}+{\dfrac
{1}{2}}\kappa^{2}\cos\varphi_{0},\\
\\
\widehat{{\mathcal{L}}}_{u}=-{\dfrac{1}{2}}\partial_{\bar{x}}^{2}+{\dfrac
{1}{2}}\kappa^{2}\cos\varphi_{0}+{\dfrac{1}{2}}\Delta(\bar{x})^{2}.
\end{array}
\right.  \label{operators}%
\end{equation}
The "gap function" reads as
\begin{align}
\Delta(\bar{x})  &  =\sqrt{2{\bar{q}}_{0}(\partial_{\bar{x}}\varphi
_{0})-(\partial_{\bar{x}}\varphi_{0})^{2}}\nonumber\\
&  =2\sqrt{{\bar{q}}_{0}\mathrm{dn\,}\left(  \bar{x},\kappa\right)
-\mathrm{dn}^{2}\mathrm{\,}\left(  \bar{x},\kappa\right)  }, \label{gapfun}%
\end{align}
where the relation $\partial_{\bar{x}}\varphi_{0}=2\mathrm{dn\,}\left(
\bar{x},\kappa\right)  $ was used. The minimum and maximum value of the gap
are given by%
\begin{equation}
\Delta_{\max}={\bar{q}}_{0},\;\;\;\Delta_{\min}=\Delta(K)=2\sqrt
{\kappa^{\prime}}\sqrt{{\bar{q}}_{0}-\kappa^{\prime}},
\end{equation}
respectively, where $\kappa^{\prime}=\sqrt{1-\kappa^{2}}$ is the complementary
modulus. We see that the gap closes at the C-IC transition. In
Fig.\ref{gapfunction} (a), we show the spatial variation of the gap function.
The $\kappa$ dependence of the minimum gap is shown in Fig.\ref{gapfunction}
(b). For small $\kappa$,$\;$we have $\Delta_{\max}\simeq2-\kappa^{2}%
/2-3\kappa^{4}/32$, and $\Delta_{\min}\simeq2-\kappa^{2}/2-7\kappa^{4}/32$.
Therefore, $\Delta_{\max}/\Delta_{\min}\simeq1$ and it is appropriate to
approximate $\Delta(\bar{x})\simeq2$ for the case of weak field. This
approximation amounts to approximating $\mathrm{dn\,}\left(  \bar{x}%
,\kappa\right)  \simeq1$.%

If were we considered only the tangential $\varphi$-mode, our problem reduces
to the case first investigated by Sutherland.\cite{Sutherland73} Furthermore,
the $\varphi$-mode is fully studied in the context of the chiral
helimagnet.\cite{Izyumov-Laptev86,Aristov-Luther03} However, to realize the
longitudinal magnetic current, as we will see, it is essential to include into
consideration the $\theta$-mode. Even of zero-field, $\kappa=0$, the $\theta
$-mode acquires the energy gap $\left(  JS^{2}\right)  \left(  m/\kappa
\right)  {\bar{q}}_{0}=DS^{2}$.\cite{BKO08} The $\theta$-gap directly
originates from the Dzyaloshinskii interaction that plays a role of easy plane
anisotropy. On the other hand, the $\varphi$-mode is the massless Goldstone
mode corresponding to rigid rotation of the whole helix around the helical
axis.\cite{Elliot66} Even after switching the perpendicular field, the
$\theta$-mode ($\varphi$-mode) remains to be massive (massless).

The mode expansion is
\begin{equation}
v(\bar{x})=\sum_{\alpha}\eta_{\alpha}v_{\alpha}(\bar{x}),\;u(\bar{x}%
)=\sum_{\alpha}\xi_{\alpha}u_{\alpha}(\bar{x}), \label{mode}%
\end{equation}
where the orthonormal basis $v_{\alpha}(\bar{x})$ and $u_{\alpha}(\bar{x})$ is
determined through the eigenvalue equations,%
\begin{equation}
\widehat{{\mathcal{L}}}_{v}v_{\alpha}(\bar{x})=\rho_{\alpha}v_{\alpha}(\bar
{x}),\;\widehat{{\mathcal{L}}}_{u}u_{\alpha}(\bar{x})=\lambda_{\alpha
}u_{\alpha}(\bar{x}), \label{eme}%
\end{equation}
with a mode index $\alpha$. The vibrational part is now given by
\begin{equation}
{\mathcal{V}}=\int_{0}^{\bar{L}}d\bar{x}\left(  \overline{{\mathcal{H}}}%
_{u}+\overline{{\mathcal{H}}}_{v}\right)  =\sum_{\alpha}\left(  \rho_{\alpha
}\eta_{\alpha}^{2}+\lambda_{\alpha}\xi_{\alpha}^{2}\right)  .
\label{vibrational_mode}%
\end{equation}
In explicit form the eigensystem (\ref{eme}) present the Schr\"{o}dinger-type
equations,
\begin{align}
{\frac{d^{2}v_{\alpha}(\bar{x})}{d\bar{x}^{2}}}  &  =[2\kappa^{2}%
\mathrm{sn\,}^{2}\left(  \bar{x},\kappa\right)  -({\kappa^{2}}+{2\rho_{\alpha
}})]v_{\alpha}(\bar{x}),\label{Lame_v}\\
{\frac{d^{2}u_{\alpha}(\bar{x})}{d\bar{x}^{2}}}  &  =[2\kappa^{2}%
\mathrm{sn\,}^{2}\left(  \bar{x},\kappa\right)  -({\kappa^{2}-}4{\bar{q}}%
_{0}+4+{2\lambda_{\alpha}})]u_{\alpha}(\bar{x}).\nonumber\\
&  \label{Lame_u}%
\end{align}
In Eq.~(\ref{Lame_u}) we consider the case of weak field corresponding to
small $\kappa$ that admit $\mathrm{dn\,}\left(  \bar{x},\kappa\right)
\simeq1$. In appendix A, we present the general scheme to treat the periodic
potential having the spatial period $2K$ and show that this approximation does
not affect qualitative result presented below. Now, both equations
(\ref{Lame_u}) and (\ref{Lame_v}) reduce to the Jacobi form of the
Lam$\acute{\mathrm{e}}$ equation,\cite{WW} and their solutions have been
discussed by us previously\cite{BKO08} (see also Appendix B). The analysis
shows that both the $\varphi$ and $\vartheta$ mode consist of two
bands,\cite{Sutherland73} i.e.,
\begin{equation}
\left\{
\begin{array}
[c]{l}%
\text{Acoustic }\varphi\text{\ mode}:\\
\;\;\;\;\;\;\omega_{\varphi}^{(-)}=\sqrt{\rho_{a}^{(-)}}=\dfrac{\kappa
^{\prime}}{\sqrt{2}}\left\vert \mathrm{sn\,}\left(  a,\kappa^{\prime}\right)
\right\vert \,,\\
\text{Optical }\varphi\text{\ mode}:\\
\;\;\;\;\;\;\omega_{\varphi}^{(+)}=\sqrt{\rho_{a}^{(+)}}=\dfrac{1}{\sqrt
{2}\left\vert \mathrm{sn\,}\left(  a,\kappa^{\prime}\right)  \right\vert },
\end{array}
\right.
\end{equation}%
\begin{equation}
\left\{
\begin{array}
[c]{l}%
\text{Acoustic }\vartheta\text{\ mode}:\ \\
\;\;\;\;\omega_{\vartheta}^{(-)}=\sqrt{\lambda_{a}^{(-)}}=\sqrt{2{\bar{q}}%
_{0}-2+\dfrac{\kappa^{\prime2}}{2}\,\mathrm{sn\,}^{2}\left(  a,\kappa^{\prime
}\right)  },\;\\
\text{Optical }\vartheta\text{\ mode}:\\
\;\;\;\;\omega_{\vartheta}^{(+)}=\sqrt{\lambda_{a}^{(+)}}=\sqrt{2{\bar{q}}%
_{0}-2+\dfrac{1}{2\mathrm{sn\,}^{2}\left(  a,\kappa^{\prime}\right)  }},
\end{array}
\right.
\end{equation}
where the real parameter $a$\ runs over $\ K^{\prime}<a\leq K^{\prime}$. Here,
$K^{\prime}$ means the complete elliptic integral of the first kind with the
complementary modulus $\kappa^{\prime}=\sqrt{1-\kappa^{2}}$.

By imposing the periodic boundary condition, the quasi-momentum (Floquet
index) is introduced for the acoustic
\begin{equation}
Q_{a}^{(-)}=\dfrac{\pi a}{2KK^{\prime}}+Z\left(  a,\kappa^{\prime}\right)  ,
\end{equation}
($0\leq|Q_{a}^{(-)}|\leq{\pi}/{2K}$) and the optic
\begin{equation}
Q_{a}^{(+)}={\dfrac{\pi a}{2KK^{\prime}}}+Z\left(  a,\kappa^{\prime}\right)
+\mathrm{dn\,}\left(  a,\kappa^{\prime}\right)  {\dfrac{\mathrm{cn\,}\left(
a,\kappa^{\prime}\right)  }{\mathrm{sn\,}\left(  a,\kappa^{\prime}\right)  }}
\,
\end{equation}
(${\pi}/{2K}\leq|Q_{a}^{(-)}|$) branches, respectively, where $Z$ denotes the
Jacobi's Zeta-function.\cite{WW}  The representation was given by Izyumov and
Laptev,\cite{Izyumov-Laptev86} and differs from a conventional
representation.\cite{WW,Sutherland73}

\begin{figure}
[h]
\begin{center}
\includegraphics[
height=3.4463in,
width=2.2399in
]%
{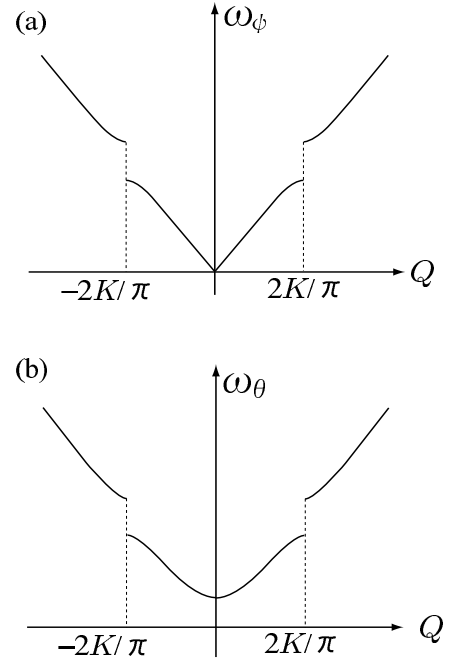}%
\caption{ The energy dispersions of the eigenmodes for (a) the tangential
$\varphi$-fluctuation ($\omega_{\varphi}$) and (b) the longitudinal $\theta
$-fluctuations ($\omega_{\theta}$). }%
\label{fig:dispersion}%
\end{center}
\end{figure}

A dispersion relation is given by $\omega$ as a function of Floquet index $Q$.
We show the excitation spectra $\omega_{\varphi}$ and $\omega_{\vartheta}$ in
Fig.~\ref{fig:dispersion}. Because $4/\pi<{\bar{q}_{0}}\leq2,$ the energy gap
of the $\vartheta$ mode, $\Delta_{\vartheta}(a=0)=\sqrt{2{\bar{q}}_{0}-2},$
has a range $\sqrt{8/\pi-2}<\Delta\leq\sqrt{2}.$ The gap has a maximum value
$\Delta_{\vartheta}=\sqrt{2}$ at zero field ($\kappa=0$) and monotonously
decreases as the field increases up to the critical field ($\kappa=1$). The
normalized wave function at the bottom of the acoustic band is%
\begin{equation}
{\Lambda}_{{\alpha=0}}(\bar{x})=\sqrt{\frac{K\left(  \kappa\right)  }{E\left(
\kappa\right)  \bar{L}}}\mathrm{dn\,}\left(  \bar{x},\kappa\right)  =\frac
{1}{2}\sqrt{\frac{K\left(  \kappa\right)  }{E\left(  \kappa\right)  \bar{L}}%
}\mathrm{\,}\partial_{\bar{x}}\varphi_{0}\left(  \bar{x}\right)  .
\label{zero_mode}%
\end{equation}
In the next section we demonstrate that ${\Lambda}_{{\alpha=0}}(\bar{x}%
)$\ exactly corresponds to the zero translational mode.%

\section{Galilean boost of the kink crystal}

In the previous section, we determined the phonon modes. Next we consider the
translational mode. The translational symmetry holds after the kink formation
and gives rise to the Goldstone mode, i.e. zero mode $\partial_{\bar{x}%
}\varphi_{0}\left(  \bar{x}\right)  =2\mathrm{dn\,}\left(  \bar{x}%
,\kappa\right)  $. Although the Gaussian fluctuations around the kink crystal
state are assumed to be small, this is not true for the zero mode which
describes fluctuations without damping. Then, the center of mass coordinate is
elevated to the status of the dynamical variable $X(t)$ and the phonon modes
are orthogonal to the zero mode. To describe this situation, we follow the
collective coordinate method.\cite{Christ-Lee75,Rajaraman}

At first, we construct the Lagrangian for the kink crystal system. We make use
of the coherent states of spins,%
\begin{equation}
\left\vert \mathbf{n}_{i}\right\rangle =\exp\left[  i\theta_{i}\mathbf{\lambda
}\cdot\mathbf{S}\right]  \left\vert S,S\right\rangle ,
\end{equation}
where
\begin{equation}
\mathbf{\lambda}={\frac{\mathbf{n}_{0}\times\mathbf{n}_{i}}{\left\vert
\mathbf{n}_{0}\times\mathbf{n}_{i}\right\vert }},
\end{equation}
with $\mathbf{n}_{i}=(\cos\theta_{i},\sin\theta_{i}\cos\varphi_{i},\sin
\theta_{i}\sin\varphi_{i})$ and $\mathbf{n}_{0}=(1,0,0).$ $S_{i}(i=x,y,z)$ are
the generators of SU(2) in the spin-$S$ representation and satisfy $\left[
S_{\alpha},S_{\beta}\right]  =i\epsilon_{\alpha\beta\gamma}S_{\gamma}.$ The
highest weight state $\left\vert S,S\right\rangle $ satisfies $S_{i}%
^{x}\left\vert S,S\right\rangle =S\left\vert S,S\right\rangle $ and
$\mathbf{S}^{2}\left\vert S,S\right\rangle =S(S+1)\left\vert S,S\right\rangle
$. The states $\left\vert \mathbf{n}_{i}\right\rangle $ form an overcomplete
set and give $\left\langle \mathbf{n}_{i}\right\vert \mathbf{S}\left\vert
\mathbf{n}_{i}\right\rangle =S\mathbf{n}_{i}.$ Using this representation, the
Berry phase contribution to the real-time Lagrangian per unit area is written
as%
\begin{align}
{\mathcal{L}}_{\text{Berry}}  &  =\hslash S\sum_{i}(\cos\theta_{i}%
-1)\partial_{t}\varphi_{i}\nonumber\\
&  =\hslash S\frac{\kappa}{m}\int_{0}^{\bar{L}}d\bar{x}(\cos\theta
-1)\partial_{t}\varphi, \label{berry}%
\end{align}
where we took the continuum limit in the second line. Now, we construct the
Lagrangian,
\begin{equation}
{\mathcal{L}}=c_{0}\int_{0}^{\bar{L}}d\bar{x}(\cos\theta-1)\partial_{t}%
\varphi-c_{1}{\mathcal{V}}, \label{Lagrangiandensity}%
\end{equation}
with the coefficients
\begin{equation}
c_{0}=\hslash S\frac{\kappa}{m},\;\;\;\;c_{1}=JS^{2}\frac{m}{\kappa},
\label{coefficients}%
\end{equation}
and expand $\varphi$\ and $\theta$\ in the form,
\begin{equation}
\left\{
\begin{array}
[c]{c}%
\varphi\left(  \bar{x},t\right)  =\varphi_{0}\left(  \bar{x}-\overline
{X}(t)\right)  +\sum_{\alpha\neq0}^{\infty}\eta_{\alpha}(t)v_{\alpha}\left(
\bar{x}-\overline{X}(t)\right)  ,\\
\theta\left(  \bar{x},t\right)  =\pi/2+\sum_{\alpha\neq0}^{\infty}\xi_{\alpha
}(t)u_{\alpha}\left(  \bar{x}-\overline{X}(t)\right)  .
\end{array}
\right.  \label{ex}%
\end{equation}
In the expansion of the $\theta$-mode, it is not necessary to exclude
$\alpha=0$, since the $\theta$-mode does not contain zero mode. This
description amounts to using the curvilinear basis, $\left\{  X,\eta_{\alpha
},\xi_{\alpha}\right\}  ,$ in functional space and taking the generalized
coordinates $q_{1}=\overline{X}$, $q_{2\alpha}=\eta_{\alpha}$, $q_{3\alpha
}=\xi_{\alpha}$. Since the zero mode $\partial_{\bar{x}}\varphi_{0}\left(
\bar{x}\right)  $ is orthogonal to the phonon modes, we have
\begin{equation}
\int_{0}^{\bar{L}}d\bar{x}\dfrac{\partial\varphi_{0}\left(  \bar{x}\right)
}{\partial\bar{x}}v_{\alpha}\left(  \bar{x}\right)  =0, \label{orthogonalilty}%
\end{equation}
for $\alpha\neq0.$ Noting that
\[
\dot{\varphi}=-\dot{q}_{1}\left(  \partial_{\bar{x}}\varphi_{0}+\sum_{\alpha
}^{\infty}q_{2\alpha}\partial_{\bar{x}}v_{\alpha}\right)  +\sum_{\alpha
}^{\infty}\dot{q}_{2\alpha}v_{\alpha},
\]
and
\[
1-\cos{\theta}\simeq1+\sum_{\alpha}^{\infty}q_{3\alpha}u_{\alpha},
\]
and plugging these expressions into the Lagrangian (\ref{Lagrangiandensity}),
we obtain%
\begin{align}
{{\mathcal{L}}}  &  =-c_{0}\left(  \sum_{\alpha}{\mathcal{J}}_{\alpha}\dot
{q}_{2\alpha}-\dot{q}_{1}\sum_{\alpha}{\mathcal{K}}_{\alpha}q_{3\alpha}\right.
\label{Lagrangian}\\
&  \;\;\;\;\;\;\;\;\;\;\;\;\;\left.  +\sum_{\alpha,\beta}{\mathcal{M}}%
_{\alpha,\beta}q_{3\alpha}\dot{q}_{2\beta}\right)  -c_{1}{\mathcal{V}},
\end{align}
where higher order terms ${\mathcal{O}}\left(  q^{3}\right)  $ are dropped.
The overlap coefficients are given by
\begin{equation}
\left\{
\begin{array}
[c]{c}%
{\mathcal{J}}_{\alpha}=\int_{0}^{\bar{L}}d\bar{x}\,v_{\alpha}\left(  \bar
{x}\right)  ,\\
{\mathcal{K}}_{\alpha}=\int_{0}^{\bar{L}}d\bar{x}\,\dfrac{\partial\varphi
_{0}\left(  \bar{x}\right)  }{\partial\bar{x}}u_{\alpha}\left(  \bar
{x}\right)  ,\\
{\mathcal{M}}_{\alpha\beta}=\int_{0}^{\bar{L}}d\bar{x}\,u_{\alpha}\left(
\bar{x}\right)  v_{\beta}\left(  \bar{x}\right)  .
\end{array}
\right.
\end{equation}

The Lagrangian (\ref{Lagrangian}) is \textit{singular} because it does not
contain any term of the form $\dot{q}_{i}\dot{q}_{j}$, and the rank of the
Hessian matrix $\left(  {\partial}^{2}{\mathcal{L}}/{\partial}\dot{q}%
_{i}{\partial}\dot{q}_{j}\right)  $ becomes zero. This means that there is no
primary expressible velocities. Therefore we need to construct the Hamiltonian
by using the Dirac's algorithm for the constrained Hamiltonian
systems.\cite{Dirac,Gitman} The details of the treatment have been given in
our previous treatment (see, also Appendix C). The final result is
\begin{equation}
\eta_{\alpha}=0, \quad\xi_{\alpha}=\frac{c_{0}}{2c_{1}}\frac{{\mathcal{K}%
}_{\alpha}}{{\lambda_{\alpha}}}\dot{\overline{X}} , \label{canonicalrelation1}%
\end{equation}
which means that \textit{only finite amplitude of the }$\vartheta
$\textit{-mode,}%
\begin{equation}
u(\bar{x})=\sum_{\alpha}\xi_{\alpha}u_{\alpha}(\bar{x}),
\end{equation}
\textit{\ appears when the collective velocity }$\dot{X}$\textit{\ is finite.}
This is exactly the manifestation of the ODLRO. In other words, the $u(\bar
{x})$-field is interpreted as the \textit{demagnetization field} that drives
the inertial motion of the kink. Using Eq.(\ref{canonicalrelation1}), we reach
the final form of the physical Hamiltonian,%
\begin{equation}
H_{\text{ph}}=c_{1}\sum_{\alpha}\lambda_{\alpha}\xi_{\alpha}^{2}=\frac
{c_{0}^{2}}{4c_{1}}\sum_{\alpha}\frac{{\mathcal{K}}_{\alpha}^{2}}%
{{\lambda_{\alpha}}}\dot{\overline{X}}^{2}=\frac{1}{2}M\dot{X}^{2},
\label{physicalHamiltonian}%
\end{equation}
where the inertial mass of the kink crystal is introduced
\begin{equation}
M=\frac{c_{0}^{2}}{2c_{1}}\left(  \frac{m}{\kappa}\right)  ^{2}\sum_{\alpha
}\frac{{\mathcal{K}}_{\alpha}^{2}}{{\lambda_{\alpha}}}. \label{mass}%
\end{equation}
The physical Hamiltonian (\ref{physicalHamiltonian}) describes the inertial
motion of the kink crystal.


The linear momentum per unit area carried by the kink crystal may be presented
in the form\cite{BKO08} $P=2\pi\hslash S{\mathcal{Q}}+M\dot{X}$, where the
topological charge
\begin{equation}
{\mathcal{Q}}=\frac{1}{2\pi}[\varphi_{0}(\bar{L})-\varphi_{0}(0)]
\label{topological_charge}%
\end{equation}
is introduced. Apparently, the transverse magnetic field increases a period of
the kink crystal lattice and diminishes the topological charge ${\mathcal{Q}}$
and therefore it affects only the\textit{\ background linear momentum} (see
discussion in Sec. VI). The physical momentum related with a mass transport
due to the excitations around the kink crystal state is generated by the
steady movement.

The \textquotedblleft superfluid magnon current\textquotedblright%
\thinspace\ transferred by the $\theta$-fluctuations is determined through the
definition of the accumulated magnon density\cite{Volovik07} $\mathcal{\rho
}_{\text{s}}$ in the total magnon density $\mathcal{N}= g \mu_{\text{B}} S
\left(  1- \cos\theta\right)  $ $=\mathcal{\rho}_{\text{0}}+\mathcal{\rho
}_{\text{s}}$, where the superfluid part $\mathcal{\rho}_{\text{s}}%
=-g\mu_{\text{B}}S\cos\theta$ is conjugated with the magnon time-even current
carried by the $\theta$-fluctuations
\begin{equation}
j^{x}(\bar{x})=g\mu_{B}S\frac{c_{0}}{2c_{1}}\frac{m}{\kappa}\,\dot{X}^{2}%
\sum_{\alpha}\frac{{\mathcal{K}}_{\alpha}}{\lambda_{\alpha}}u_{\alpha}\left(
\bar{x}\right)  \label{AcSpCur}%
\end{equation}
via a continuity equation.\cite{BKO08} The important point is that the only
massive $\theta$-mode can carry the longitudinal magnon current as a
manifestation of ordering in non-equilibrium state, i.e., dynamical
off-diagonal long range order.\cite{Xiao}

The net magnetization (magnetic dipole moment) induced by the movement is
\begin{align}
m(\bar{x})  &  \simeq-g\mu_{B}S{u}(\bar{x})\nonumber\\
&  =-g\mu_{B}S\frac{c_{0}}{2c_{1}}\frac{m}{\kappa}\,\dot{X}\sum_{\alpha}%
\frac{{\mathcal{K}}_{\alpha}}{\lambda_{\alpha}}u_{\alpha}\left(  \bar
{x}\right)  . \label{NetMag}%
\end{align}
The minus sign means that the net magnetization produces a demagnetization field.

\section{Quantitative Estimates}

{To compute the mass $M$, the spin current $j^{x}$ and the magnetic dipole
moment $m$, we consider an array of parallel chains described by the model
(\ref{lattH}), where a number of chains \textit{per unit area} is
$n_{\text{area}}=1/a_{0}^{2}$. In the case of the molecular-based chiral
magnets, the crystal packing is usually loose ($a_{0}\simeq10^{-9}$[m]) and
the exchange interaction is rather weak ($J\simeq10$[K] $\simeq10^{-22}$[J]).
On the other hand, in the case of the inorganic chiral magnets, the crystal
packing is close ($a_{0}\simeq10^{-10}$[m]) and the exchange interaction is
rather strong ($J\simeq100$[K] $\simeq10^{-21}$[J]). We take these values as
just typical parameter choices. The strength of the Dzyaloshinskii interaction
is ambiguous and we simply take $q_{0}=D/J=10^{-2}$.}

\subsection{Mass of the kink crystal}

The mass $M$ of the kink crystal is given by Eq.(\ref{mass}). Evaluation of
the overlap integral ${\mathcal{K}}_{\alpha}$ is performed in appendix D and
yields ${\mathcal{K}}_{\alpha}=\delta_{\alpha,0}{\mathcal{K}}_{0}$, where
\begin{equation}
{\mathcal{K}}_{0}=2\sqrt{\frac{E\left(  \kappa\right)  }{K\left(
\kappa\right)  }\dfrac{m}{\kappa}\frac{L}{a_{0}}}. \label{overlap}%
\end{equation}
Therefore we have
\begin{equation}
M=\frac{c_{0}^{2}}{2c_{1}}\left(  \dfrac{m}{\kappa}\right)  ^{2}%
\frac{{\mathcal{K}}_{0}^{2}}{\lambda_{0}}\frac{1}{a_{0}^{2}}.
\end{equation}
The factor $1/a_{0}^{2}$ appears here after the MKS units [m] for distances
are recovered in Eq.(\ref{physicalHamiltonian}). The mass per unit area is
given by
\begin{equation}
M_{\text{area}}=n_{\text{area}}\times M=\frac{c_{0}^{2}}{2c_{1}}\left(
\dfrac{m}{\kappa}\right)  ^{2}\frac{{\mathcal{K}}_{0}^{2}}{\lambda_{0}}%
\frac{1}{a_{0}^{4}}, \label{massa0}%
\end{equation}
that after simplification yields
\[
M_{\text{area}}=\frac{2E\left(  \kappa\right)  }{\lambda_{0}K\left(
\kappa\right)  }\frac{\hslash^{2}L}{Ja_{0}^{5}}\simeq\frac{\hslash^{2}%
L}{Ja_{0}^{5}}.
\]
The last relationship is reliable in the case of small fields, i.e.
$\lambda_{0}\simeq2$, and $K\left(  \kappa\right)  =E\left(  \kappa\right)
\simeq\pi/2$.

Noting that the period of kink measured in lattice units is given by
Eq.(\ref{period}), which turns into $\ell_{\text{kink}}={8K(\kappa)}%
E(\kappa)/\pi q_{0}\simeq2\pi J/D$ for small fields, the mass per one kink
acquires the form
\begin{equation}
M_{\text{kink}}=M_{\text{area}}\frac{\ell_{\text{kink}}}{L}\simeq\frac{J}%
{D}\frac{\hslash^{2}}{Ja_{0}^{4}}.
\end{equation}
{As a typical example of the molecular-based chiral magnets, we have
\[
M_{\text{kink}}\simeq10^{-9}[\text{g/cm}^{2}].
\]
For the chain length $L/a_{0}=10^{5}$, we have the total mass }$M_{\text{area}%
}${$\simeq\frac{D}{J}\frac{L}{a_{0}}M_{\text{kink}}{\simeq}10^{-4}$[g/cm$^{2}%
$]. As a typical example of the inorganic chiral magnets, we have
\[
M_{\text{kink}}\simeq10^{-6}[\text{g/cm}^{2}].
\]
For the chain length $L/a_{0}=10^{6}$, we have the total mass }$M_{\text{area}%
}${$\simeq10^{-2}$[g/cm$^{2}$]. This heavy mass should be compared with the
mass of conventional Bloch wall mass in ferromagnets. To make clear the
difference, in appendix E, we gave a brief summary of this issue. In the
present case, appearance of the heavy mass is easily understood, since the
kink crystal consists of a macroscopic array of large numbers of local kinks.
}

\subsection{Spin current}

\begin{figure}
[h]
\begin{center}
\includegraphics[
height=0.7342in,
width=3.0761in
]%
{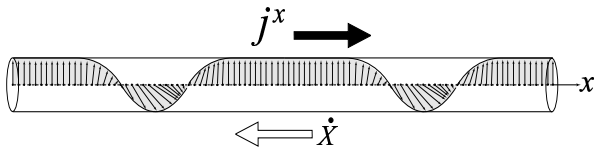}%
\caption{A schematic view of an instant distribution of spins in the
current-carrying state. This picture corresponds to the case of intermediate
field strength. }%
\label{Volovik}%
\end{center}
\end{figure}

\begin{figure}
[h]
\begin{center}
\includegraphics[
height=1.7504in,
width=3.1687in
]%
{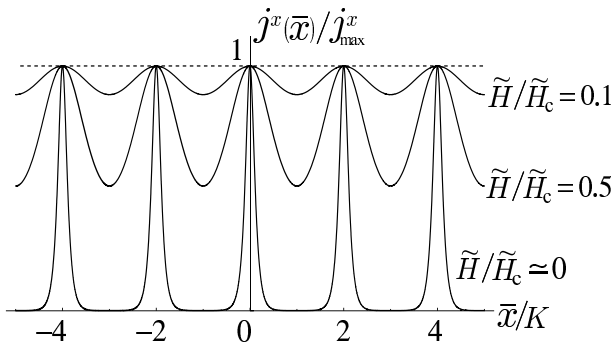}%
\caption{ A snapshot of the position dependence of the current density
$j^{x}(\tilde{x})$. $j^{x}(\tilde{x})$ is scaled by its maximum
$j_{\mathrm{max}}^{x}=j^{x}(0)$. We depicted the cases of the magnetic field
strengths $\tilde{H}/\tilde{H}_{c}=0.1$, $0.5$, and $\tilde{H}/\tilde{H}%
_{c}\simeq1$. }%
\label{fig:current}%
\end{center}
\end{figure}

As it follows from Eq.(\ref{AcSpCur}) the physical dimension of the spin
current density is $\text{Wb}\cdot{\text{m}}^{2}/{\text{s}}$. Using the
results of the appendix D the spin current density given by Eq.(\ref{AcSpCur})
transforms into
\begin{equation}
j^{x}(\bar{x})=g\mu_{B}S\frac{c_{0}}{2c_{1}}\frac{m}{\kappa}\frac{1}{a_{0}%
}\,\dot{X}^{2}\,\frac{{\mathcal{K}}_{0}}{\lambda_{0}}u_{0}\left(  \bar
{x}\right)  .
\end{equation}
The factor $1/a_{0}$ occurs after the MKS units for distances are recovered in
the continuity equation $\partial_{x}\rightarrow a_{0}\partial_{x}$ and in the
velocity $\dot{X}\rightarrow\dot{X}/a_{0}$. After simplifications with aid of
Eqs.(\ref{period}), (\ref{coefficients}), (\ref{zero_mode}), and
(\ref{overlap}) we immediately have
\begin{equation}
j^{x}(\bar{x})=\frac{g\mu_{B}\hbar}{Ja_{0}}\frac{4E\left(  \kappa\right)
}{\pi q_{0}}\frac{1}{\lambda_{0}}\,\dot{X}^{2}\,\mathrm{dn\,}\left(  \bar
{x},\kappa\right)  . \label{currentformula}%
\end{equation}
For the case of weak fields corresponding to small $\kappa$ this yields
\begin{equation}
j^{x}(\bar{x})\simeq\frac{g\mu_{B}\hbar}{Ja_{0}q_{0}}\,\dot{X}^{2}%
\,\mathrm{dn\,}\left(  \bar{x},\kappa\right)  . \label{current2}%
\end{equation}
{We present a schematic view of an instant distribution of spins in the
current-carrying state in Fig.~\ref{Volovik}. In Fig.~\ref{fig:current}, we
present a snapshot of the position dependence of the current density
$j^{x}(x)$ in the weak field limit, given by Eq.~(\ref{current2}). In
Fig.~\ref{fig:current}, we depicted the cases of the magnetic field strengths
$\tilde{H}/\tilde{H}_{c}=0.1$, $0.5$, and $\tilde{H}/\tilde{H}_{c}\simeq1$.
Although the formula (\ref{current2}) is valid only for the case of weak field
limit, but qualitative features are well demonstrated by just extrapolating
the validity up to $\tilde{H}/\tilde{H}_{c}\simeq1$. As the field strength
approaches the critical value, the current density is more and more localized.
}

{For both the typical molecular-based and inorganic chiral magnets, we have
\[
j^{x}(\bar{x})\sim0.1\,\mu_{B}\,\dot{X}^{2}\sim10^{-24}\,\dot{X}%
^{2}\,[\text{Wb}\cdot{\text{s}}].
\]
Taking the velocity of order $\dot{X}\sim10^{2}\,[\text{m/s}]$ we obtain
finally
\[
j^{x}(\bar{x})\sim10^{-20}\,[\text{Wb}\cdot{\text{m}}^{2}/{\text{s}}],
\]
therefore the current through the unit area
\[
j_{\text{area}}^{x}(\bar{x})=j^{x}(\bar{x})\times n_{\text{area}}%
\sim1\,[\text{Wb}/{\text{s}}].
\]
}%

\subsection{Magnetic {dipole moment}}

The magnetic {dipole moment} [Eq.(\ref{NetMag})] induced by the motion is
given by
\[
m(\bar{x})=-g\mu_{B}S\frac{c_{0}}{2c_{1}}\frac{m}{\kappa}\frac{1}{a_{0}}%
\,\dot{X}\,\frac{{\mathcal{K}}_{0}}{\lambda_{0}}u_{0}\left(  \bar{x}\right)
,
\]
i.e. the relationship $j^{x}=-m\,\dot{X}$ holds. By the same manner as it was
made for the spin current we obtain in the case of the small fields
\begin{equation}
m(\bar{x})\simeq-\frac{g\mu_{B}\hbar}{Ja_{0}q_{0}}\,\dot{X}\,\mathrm{dn\,}%
\left(  \bar{x},\kappa\right)  .
\end{equation}
{Therefore, for both the molecular-based and inorganic chiral magnets, we
have
\begin{equation}
m(\bar{x})\sim0.1\,\mu_{B}\,\dot{X}\sim10\,\mu_{B},
\end{equation}
i.e. $m(\bar{x})$ is of order $10^{-22}\,[\text{Wb}\cdot{\text{m}}]$.} The
total magnetic moment of the chain is
\begin{align}
m_{\text{chain}}  &  \simeq-\frac{g\mu_{B}\hbar}{Ja_{0}q_{0}}\,\dot{X}%
\,\int_{0}^{\bar{L}}\mathrm{dn\,}\left(  \bar{x},\kappa\right)  \,d\bar
{x}\nonumber\\
&  =-\frac{g\mu_{B}\hbar}{Ja_{0}q_{0}}\,\pi{\mathcal{Q}}\dot{X}.
\end{align}
We here used the relations $\int\mathrm{dn\,}\left(  \bar{x},\kappa\right)
\mathrm{\,}d\bar{x}=\mathrm{am\,}\left(  \bar{x},\kappa\right)  $ and
$[\varphi_{0}(x)+\pi]/2=\sin^{-1}\left[  \mathrm{sn\,}\left(  \bar{x}%
,\kappa\right)  \right]  =\mathrm{am\,}\left(  \bar{x},\kappa\right)  $ that
leads to
\begin{equation}
\int_{0}^{\bar{L}}\mathrm{dn\,}\left(  \bar{x},\kappa\right)  \mathrm{\,}%
d\bar{x}=\frac{1}{2}[\varphi_{0}(\bar{L})-\varphi_{0}(0)]=\pi{\mathcal{Q}},
\end{equation}
where ${\mathcal{Q}}$\ is a topological charge introduced in
Eq.(\ref{topological_charge}).

Noting,
\begin{equation}
{\mathcal{Q}}=L/l_{\text{kink}}=\frac{\pi q_{0}L}{8K\left(  \kappa\right)
E\left(  \kappa\right)  a_{0}},
\end{equation}
we have the chain magnetization
\[
m_{\text{chain}}\simeq-\frac{g\mu_{B}\hbar}{2Ja_{0}}\left(  \frac{L}{a_{0}%
}\right)  \,\dot{X}.
\]
The total moment per unit volume
\[
m_{\text{vol}}=m_{\text{chain}}\times n_{\text{area}}\times L^{2}\simeq
-\frac{g\mu_{B}\hbar}{2Ja_{0}}\left(  \frac{L}{a_{0}}\right)  ^{3}\,\dot{X}.
\]
{As a typical example of the molecular-based chiral magnets, we have
\[
m_{\text{vol}}\sim10^{-11}\,\dot{X}\sim10^{-9}[\text{Wb}\cdot\text{m}].
\]
As a typical example of the inorganic chiral magnets, we have
\[
m_{\text{vol}}\sim10^{-8}\,\dot{X}\sim10^{-6}[\text{Wb}\cdot\text{m}].
\]
}

\section{Discussions of related topics}

\subsection{Background spin current problem: SU(2) gauge invariant
formulation}

Heurich, K\"{o}nig and MacDonald{\cite{Heurich03}} proposed that the external
magnetic fields generate dissipationless spin currents in the ground state of
systems with spiral magnetic order. Here, we comment on the relevance of the
present work to this issue. In our model, the \textit{background }spin current
is given by
\begin{equation}
j_{\text{bg}}={\partial\varphi_{0}(\bar{x})}/{\partial\bar{x}}-{\bar{q}}%
_{0}\propto\mathrm{dn\,}(\bar{x})-2E(\kappa)/\pi,
\end{equation}
i.e. there arises the misfit of the kink crystal to the helimagnetic
modulation and consequently the current comes up. Below we prove that this
current exists on a link between two sites but it causes no accumulation of
magnon density ("magnetic charge") at the site due to continuity equation,
i.e. the current is not related to the magnon transport. This supports
reasonings of arguments by Sch\"{u}tz, Kopietz, and M.~Kollar{\cite{Kollar}}
that appearance of finite spin currents is direct manifestation of quantum
correlations in the system, and in the classical ground state the spin
currents vanish.

The background spin current problem is best described by the SU(2) gauge
invariant formulation developed by Chandra, Coleman and Larkin.\cite{CCL90} By
imposing the local SU(2) gauge invariance of the theory, we obtain the
fictitious SU(2) gauge fields $\mathbf{a}$ and $\mathbf{h}$ that give the spin
current ${\mathbf{J}}^{(S)}=\partial\mathcal{L}_{g}/\partial{\mathbf{a}}$, and
the spin density ${\mathbf{S=}}\partial\mathcal{L}_{g}/\partial{\mathbf{h}}$,
respectively, where $\mathcal{L}_{g}$ is the gauge-invariant Lagrangian.

Following Chandra, Coleman and Larkin, we use the SU(2) Schwinger boson
representation,%
\begin{equation}
{\mathbf{S}}_{i}=\frac{1}{2}b_{i\alpha}^{\dagger}{\mathbf{\sigma}}%
_{\alpha\beta}b_{i\beta},\ \ \ \sum\limits_{\alpha}b_{i\alpha}^{\dagger
}b_{i\alpha}=2S,\;\left(  \alpha=1,2\right)  \label{constr}%
\end{equation}
where ${\mathbf{\sigma}}=\left(  \sigma_{x},\sigma_{y},\sigma_{z}\right)  $
are the Pauli matrices. In the path-integral prescription, the partition
function is represented as%
\begin{equation}
\mathcal{Z}=\int\mathcal{D}b_{i\alpha}^{\dagger}\mathcal{D}b_{i\alpha
}\mathcal{D}{\lambda}_{i}\exp\left(  -\int_{0}^{\beta}\mathcal{L}(\tau
)d\tau\right)  , \label{partfun}%
\end{equation}
where the Lagrangian is given by
\begin{align}
\mathcal{L}(\tau)  &  =\sum\limits_{i}\left[  b_{i\alpha}^{\dagger}%
\partial_{\tau}b_{i\alpha}+i{\lambda}_{i}\left(  b_{i\alpha}^{\dagger
}b_{i\alpha}-2S\right)  \right] \nonumber\\
&  +\mathcal{H}\left[  S\left(  b^{\dagger},b\right)  \right]  ,
\label{Lagrang}%
\end{align}
where $\mathcal{H}$ is the Hamiltonian (\ref{lattH}) written in terms of the
Schwinger bosons, and $\tau$ represents the imaginary time. The Lagrange
multiplier ${\lambda}_{i}$ provides the local constraint. The local SU(2)
gauge transformation acting on the SU(2) doublet, $b_{i}^{+}=\left(
b_{i1}^{+},b_{i2}^{+}\right)  $ is given by
\begin{equation}
b_{i}^{\prime+}=b_{i}^{+}{\hat{g}}_{i}^{-1},\;\ \ \ b_{i}^{\prime}={\hat{g}%
}_{i}b_{i},
\end{equation}
where
\begin{equation}
\hat{g}_{i}(t)=\exp\left[  -\frac{i}{2}{\mathbf{\Theta}}_{i}(t)\cdot
\mathbf{\sigma}\right]  .
\end{equation}
The SU(2) rotation $\hat{g}_{i}$\ gives the rotation of the spin vector,%
\begin{equation}
\mathbf{S}_{i}^{\prime}=\exp\left(  -{\mathbf{\Theta}}_{i}\cdot{\mathbf{\hat
{I}}}\right)  {\mathbf{S}}_{i}\simeq{\mathbf{S}}_{i}-{\mathbf{\Theta}}%
_{i}\times{\mathbf{S}}_{i},
\end{equation}
where $(\hat{I}_{\mu})_{\nu\lambda}=\varepsilon_{\mu\nu\lambda}$ ($\mu
,\nu,\lambda=x,y,z$) is the adjoint representation of the Lie algebra of the
SO$(3)$ group characterized by $[\hat{I}_{\mu},\hat{I}_{\nu}]=\varepsilon
_{\mu\nu\lambda}\hat{I}_{\lambda}$.

Rewriting the Lagrangian in the gauge invariant form, there appears a term%
\begin{equation}
b_{i\alpha}^{\prime\dagger}\left(  {\hat{g}}_{i}\partial_{\tau}{\hat{g}}%
_{i}^{-1}\right)  b_{i\alpha}^{\prime}=ie^{{\mathbf{\Theta}}_{i}%
\cdot{\mathbf{\hat{I}}}}\,\partial_{\tau}{\mathbf{\Theta}}_{i}\cdot
{\mathbf{S}}_{i}^{\prime},
\end{equation}
that leads to introducing the gauge field ${\mathbf{h}}_{i}$ transformed as%

\begin{equation}
{\mathbf{h}}_{i}\rightarrow{\mathbf{h}}_{i}^{\prime}=e^{{\mathbf{\Theta}}%
_{i}\cdot{\mathbf{I}}}\,\left(  {\mathbf{h}}_{i}+\partial_{t}{\mathbf{\Theta}%
}_{i}\right)  , \label{Bgauge}%
\end{equation}
where $\tau=it$. Introducing the gauge covariant time derivative,
$\mathcal{D}_{t}\equiv\partial_{t}-{\mathbf{h}}\times,$we have ${\mathbf{h}%
}_{i}^{^{\prime}}={\mathbf{h}}_{i}+\mathcal{D}_{t}{\mathbf{\Theta}}_{i}.$The
fictitious magnetic field $\nabla_{t}{\mathbf{\Theta}}_{i}$ is induced by the
time-dependent rotation of the spin reference frame.

The exchange terms are regrouped in a gauge-invariant form,%

\begin{align}
&  -J\sum_{<i,j>}\mathbf{S}_{i}\cdot\mathbf{S}_{j}+\mathbf{D}\cdot\sum
_{<i,j>}\mathbf{S}_{i}\times\mathbf{S}_{j}\nonumber\\
&  =-\mathcal{J}\sum_{<i,j>}{\mathbf{S}}_{j}\,\exp\left[  -\left(  \int
_{x_{i}}^{x_{j}}{\mathbf{a}}\cdot d\mathbf{r}\right)  \ \hat{I}_{x}\right]
\,{\mathbf{S}}_{i}, \label{gaugeinv1}%
\end{align}
where $\mathcal{J=}\sqrt{J^{2}+D^{2}},$\ $x_{i}$\ represents the position of
the $i$-th site,\ and the spin vector potential is introduced as
$a_{x}=\left(  {D}/{J}\right)  \hat{\mathbf{e}}_{x}$, $a_{y}=a_{z}=0$,
corresponding to the model (\ref{lattH}). The form of Eq.(\ref{gaugeinv1})
indicates that the tangential phase angle $\varphi_{i}$\ can be gauged away by
the local rotation of the angle $\left(  D/J\right)  R_{xi}$ around the
$x$\ axis. The gauge field ${\mathbf{a}}$\ is transformed as%
\begin{equation}
{\mathbf{a}}_{i}\rightarrow{\mathbf{a}}_{i}^{\prime}=e^{{\mathbf{\Theta}}%
_{i}{\mathbf{I}}}\left(  {\mathbf{a}}-\partial_{x_{i}}{\mathbf{\Theta}}%
_{i}\right)  , \label{Agauge}%
\end{equation}
or ${\mathbf{a}}_{i}^{\prime}={\mathbf{a}}-\nabla_{x_{i}}{\mathbf{\Theta}}%
_{i}$ via the gauge covariant space derivative $\nabla_{x_{i}}\equiv
\partial_{x_{i}}+{\mathbf{a}}\times$. In addition to the physical gauge field,
$\left(  {D}/{J}\right)  \hat{\mathbf{e}}_{x}$, there appears the fictitious
gauge field, $\nabla_{x_{i}}{\mathbf{\Theta}}_{i}$, induced by the spatial
rotation of the spin reference frame.

The variation of the partition function under a local gauge transformation
must be zero \begin{widetext}
\begin{eqnarray}
\delta {\cal Z}=\int {\cal D}b_{i \alpha}^{\dagger } {\cal D}b_{i \alpha}{\cal D}{\tilde \lambda}_i \exp \left( -\int
{\cal L}_g (\tau )d\tau \right) \left( \frac{\partial {\cal L}_g}{\partial {\mathbf{a}}_{i
}^{\prime }}\cdot \delta {\mathbf{a}}_{i }^{\prime }+\frac{\partial {\cal L}_g}{\partial
{\mathbf{h}}_i^{\prime }}\cdot\delta {\mathbf{h}}_i^{\prime }\right) =0,
\end{eqnarray}
\end{widetext}where $\delta{\mathbf{a}}_{i\alpha}^{\prime}=-\nabla_{x_{i}%
}\delta{\mathbf{\Theta}}_{i},$ $\delta{\mathbf{h}}_{i}^{\prime}=\nabla
_{t}\delta{\mathbf{\Theta}}_{i}$. Consequently, one obtains the conservation
law
\begin{equation}
\nabla_{x_{i}}\left(  \partial\mathcal{L}_{g}/\partial{\mathbf{a}}_{i}%
^{\prime}\right)  -\nabla_{t}\left(  \partial\mathcal{L}_{g}/\partial
{\mathbf{h}}_{i}^{\prime}\right)  =0.
\end{equation}
By definition $\left.  \left(  \partial\mathcal{L}_{g}/\partial{\mathbf{a}%
}_{i}^{\prime}\right)  \right\vert _{{\mathbf{a}}_{i}^{\prime}={\mathbf{a}}%
}={\mathbf{J}}_{i}^{(S)}$ is the spin current, where the gauge field is fixed
by the Dzyaloshinskii vector. On the other hand $\partial\mathcal{L}%
_{g}/\partial{\mathbf{h}}_{i}^{\prime}=-{\mathbf{h}}_{i}$, and we finally
obtain the continuity equation
\begin{equation}
\nabla_{x_{i}}{\mathbf{J}}_{i}^{(S)}+\nabla_{t}{\mathbf{S}}_{i}=0, \label{CL1}%
\end{equation}
where ${\mathbf{J}}_{i}^{(S)}={\mathbf{J}}_{i\rightarrow i+1}^{(S)}%
+{\mathbf{J}}_{i-1\rightarrow i}^{(S)}$. In the explicit form, the spin
current from the site $i$ to $i+1$ is given by
\begin{align*}
{\mathbf{J}}_{i\rightarrow i+1}^{(S)}  &  =Jx_{i}\left(  {\mathbf{S}}%
_{i}\times{\mathbf{S}}_{i+1}\right)  +x_{i}\left[  \left(  {{\mathbf{D}}%
}\times{\mathbf{S}}_{i+1}\right)  \times{\mathbf{S}}_{i}\right] \\
&  =S^{2}\mathcal{J}\sin\left(  \varphi_{i+1}-\varphi_{i}-\varphi_{0}\right)
\hat{\mathbf{e}}_{x}.
\end{align*}
For the long-period incommensurate structure (${D}/{J}\leq1$) this yields in
the continuum limit
\begin{equation}
{\mathbf{J}}_{i\rightarrow i+1}^{(S)}\simeq{J}S^{2}\left(  \frac
{\partial\varphi}{\partial x}-\frac{{D}}{{J}}\right)  \hat{\mathbf{e}}_{x}.
\label{MAC}%
\end{equation}
The spin current from the site $i-1$ to the site $i$
\[
{\mathbf{J}}_{i-1\rightarrow i}^{(S)}=Jx_{i}\left(  {\mathbf{S}}_{i}%
\times{\mathbf{S}}_{i-1}\right)  -x_{i}\left[  \left(  {{\mathbf{D}}}%
\times{\mathbf{S}}_{i}\right)  \times{\mathbf{S}}_{i-1}\right]
\]
gives $-{\mathbf{J}}_{i\rightarrow i+1}^{(S)}$ in the continuum limit and
compensates (\ref{MAC}). Thus, the spin current through the $i$-th site causes
no accumulation of magnon density at the site, i.e. \textit{the current is not
transport one}. The accumulation of magnon density means that the local
quantization axis is wobbling. This wobbling motion, however, contradicts the
spontaneous symmetry breaking in the \textit{ground state}.

\subsection{Spin supercurrent in $^{3}$He}

The moving kink crystal belongs to a class of dynamical systems out of
equilibrium.\cite{Volovik07} In contrast to a class of equilibrium macroscopic
ordered state with a broken symmetry (ordered magnets, liquid crystals,
superfluids and superconductors) an emerging steady state is supported by
pumping of energy. The coherent spin precession discovered in superfluid
$^{3}$He known as homogeneously precessing domains (HPD) is a striking example
of the quantum state.\cite{Fomin91}

The precession of magnetization (spin) occurs after the magnetization is
deflected by a finite angle by the rf field from its equilibrium value. The
Larmor precession spontaneously acquires a coherent phase throughout the whole
sample. This is equivalent to the appearance of a coherent superfluid Bose
condensate, i.e. HPD is the Bose-condensate of magnons. According to the
analogy the deviation of the spin projection from its equilibrium value in the
precession plays the role of the number density of magnons. In terms of magnon
condensation the precession can be viewed as the off-diagonal long-range order
for magnons, where the phase of precession plays the role of the phase of the
superfluid order parameter, and the precession frequency plays the role of
chemical potential.

The remarkable property of the magnon Bose condensate in $^{3}$He-B is that
non-equilibrium precession has a fixed density of Bose condensate. The density
cannot relax continuously, a decay of the condensate occurs due to decreasing
volume of the superfluid part. This results in the formation of two regions of
precession: the domain with HPD is separated by a phase boundary, where a
precession frequency equals to the Larmor frequency, from the domain with
static equilibrium magnetization (non-precessing domain, NPD). In the absence
of a continuous pumping, i.e. rf field, HPD remains in the fully coherent Bose
condensate state, while the phase boundary between HPD and NPD slowly moves up
to decrease a volume of the Bose condensate.

We may suggest that in the total analogy with the supercurrents in $^{3}$He,
i.e. spin currents transferred by the coherent spin precession, the pumping of
magnons in the kink crystal (by ultrasound, for example) will cause an
appearance of homogeneously moving domains with ODLRO separated by a phase
boundary from the domain with a static soliton lattice. Without an external
flux of energy, the relaxation will occurs via gradual decrease of the volume
of the superfluid phase.

\subsection{Experimental aspects}

In realizing the bulk magnetic current proposed here, a single crystal of
chiral magnets serves as spintronics device. The mechanism involves no
spin-orbit coupling and the effect is not hindered by dephasing. Finally, we
propose possible experimental methods to trigger off the spin current
considered here.

\subsubsection{Spin torque mechanism and spin current amplification}

The spin-polarized electric current can exert torque to ferromagnetic moments
through direct transfer of spin angular momentum.\cite{Slonczewski96} This
effect, related with Aharonov-Stern effect \cite{Aharonov-Stern} for a
classical motion of magnetic moment in an inhomogeneous magnetic field, is
eligible to excite the sliding motion of the kink crystal by injecting the
spin-polarized current (polarized electron beam) in the direction either
perpendicular or oblique to the chiral axis. The spin current transported by
the soliton lattice may amplify the spin current of the injected carriers.

\subsubsection{XMCD}

To detect the magnetic dipole moment dynamically induced by the kink crystal
motion, x-ray magnetic circular dichroism (XMCD) may be used. Photon angular
momentum may be aligned either parallel or anti-parallel to the direction of
the longitudinal net magnetization.

\subsubsection{Ultrasound attenuation under the magnetic field}

Further possibility to control and detect the spin current is using a coupling
between spins and chiral torsion. Fedorov \textit{et al}\cite{Fedorov} first
pointed out that under the external torsion, the magneto-elastic coupling of
the form, $\sum_{\mathbf{R}_{i},\mathbf{R}_{j}}g_{ij}\left[  \mathbf{\nabla
}\times\left(  \mathbf{u}_{i}-\mathbf{u}_{j}\right)  \right]  \cdot
\mathbf{S}_{i}\times\mathbf{S}_{j},$ appears, where $\mathbf{u}_{i}$ is the
displacement of the magnetic atom at a lattice point $\mathbf{R}_{i}$. Then,
the quantity $\mathbf{d}_{ij}=g_{ij}\left[  \mathbf{\nabla}\times\left(
\mathbf{u}_{i}-\mathbf{u}_{j}\right)  \right]  $ plays a role of an effective
Dzyaloshinskii interaction. Ultrasound with the wavelength being adjusted to
the period of the kink crystal may resonantly modulate $\mathbf{d}_{ij}$ and
may exert the periodic torque on the kink crystal. Consequently, the kinetic
energy is supplied to the kink crystal and the ultrasound attenuation may
occur.\cite{Hu} {Then, the attenuation rate should change upon changing the
applied magnetic field strength. }

\subsubsection{TOF technique}

The most direct way of detecting the traveling magnon density may be winding a
sample by a pick-up coil and performing the time-of-flight (TOF) experiment.
Then, the coil should detect a periodic signal induced by the magnetic current.

\subsubsection{Energy loss of the moving kink crystals}

{The moving kink crystal produces the time-varying vector potential
\textit{per kink},}%
\begin{equation}
\mathbf{A}\left(  \mathbf{r},t\right)  \mathbf{=}\frac{\mu_{0}}{4\pi}%
\frac{\mathbf{m}\left(  x-Vt\right)  \times\mathbf{r}}{r^{3}},
\end{equation}
where $V=\dot{X}$ and $\mathbf{r}$\ is the position vector with respect to the
kink center. {Then, the magnitude of the induced azimuthal electric field
}$\mathbf{E}$ around the chiral axis is given by
\begin{equation}
E_{\varphi}(\rho)=\left\vert \frac{\partial A_{\varphi}}{\partial
t}\right\vert \simeq\frac{3\mu_{0}}{4\pi}\frac{g\mu_{B}\hbar}{Ja_{0}q_{0}%
}\,V^{2}{\frac{\rho x}{\left(  x^{2}+\rho^{2}\right)  ^{5/2}}}.
\end{equation}
{\ }where $\rho$\ is the radial coordinate. {Then, in the \textit{metallic}
chiral magnets, strong energy loss may occur due to the induced eddy currents.
This phenomena is exactly analogous to a well known fact that a magnet moving
through inside of the metallic pipe feels strong friction. On the other hand,
in the \textit{insulating} chiral magnets, there is no eddy current loss and
instead the polariton excitations are expected to occur. Therefore, the
frictional force acting on the moving crystal can be strongly diminished in
the insulating magnets.\cite{Chikazumi}}

\section{Concluding remarks}

In this paper, we gave a detailed account of a mechanism of possible
longitudinal transport spin current in the chiral helimagnet under transverse
magnetic field. The most important notion is that the \textquotedblleft spin
phase\textquotedblright directly comes up in the observable effects through
the soliton lattice formation. In our mechanism, the current is carried by the
moving magnetic kink crystal, where the linear momentum has a form, $P=2\pi
S{\mathcal{Q}}+M\dot{X}$. The topological magnetic charge, $S{\mathcal{Q}}$,
merely enters the equilibrium background momentum $2\pi S{\mathcal{Q}}$, while
the collective translation of the kinks with the velocity $\dot{X}$ gives the
mass $M$. Among the Gaussian fluctuations around the kink crystal state in the
soliton sector the longitudinal (along with the helical axis) $\theta$
fluctuations  play a crucial role to determine the mass of kinks. Appearance
of the spin currents is a manifestation of ordering in non-equilibrium state,
i.e., dynamical off-diagonal long range order.

We also stressed that if we took account of only the $\varphi$-fluctuations,
the spin current (Josephson current)\ would cause no accumulation of magnon
density and the current is not transport one. The accumulation of magnon
density means that the local quantization axis is wobbling but this
contradicts the spontaneous symmetry breaking in the ground state.

This mechanism is quite analogous to the D\"{o}ring-Becker-Kittel mechanism of
the domain wall motion, i.e., the Galilean boost of the solitonic kink. In our
case, the coherent motion of the kink crystal is dynamically induced by
spontaneous emergence\ of the demagnetization field. To describe the kink
crystal motion and resultant emergence of the demagnetization field, we
revisited the Sutherland's seminal work\cite{Sutherland73} and generalized it
to the case of vectorial degrees of freedom, i.e., not only the tangential
$\varphi$ but and the longitudinal $\theta$ degrees of freedom are considered.
To clarify the physical meaning of the inertial mass, we used the canonical
formulation of the kink crystal motion. We showed that in the case of
molecular-based chiral magnets, the inertial mass per kink amounts to
$M_{\text{kink}}\simeq10^{-9}[\text{g/cm}^{2}]$ and the total mass
$M_{\text{area}}\simeq10^{-4}$[g/cm$^{2}$]. In the case of the inorganic
chiral magnets, $M_{\text{kink}}\simeq10^{-6}[\text{g/cm}^{2}],$ and the total
mass $M_{\text{area}}\simeq10^{-2}$[g/cm$^{2}$]. Furthermore, the magnetic
dipole moment per kink, induced by the kink crystal motion, amounts to
$m\sim0.1\,\mu_{B}\,\dot{X}\sim10\,\mu_{B}$. {Appearance of the heavy mass is
a consequence of the fact that the kink crystal consists of a macroscopic
array of large numbers of local kinks. }

{We here mention that in our scheme, the energy gap of the} $\theta$-mode
plays a role of "protector" of the rigid sliding motion of the kink crystal.
To excite the $\theta$-mode, we need to supply the energy via the external
force. This situation is reminiscent of the existence of a threshold like
Larmor frequency in the superfluid $^{3}$He. To make clear the physical nature
of the edge velocity in our scheme is beyond the scope of the present work. We
leave this problem for future consideration.

Detection of these observable quantities may be quite a promising challenge
for experimentalists. Behind this issue, there is an actively argued problems
on how to make use of the indirect couplings among the magnetic, electronic,
and elastic degrees of freedom. For example, magnetic-field-dependent
ultrasonic attenuation may give us a new insights. To materialize the
theoretical model presented here, symmetry-adapted material synthesis would be
required. So far, a novel category of materials suitable for chiral magnets
has been successfully fabricated on purpose for application in the field of
both molecule-based and inorganic magnetic materials. The interplay of
crystallographic and magnetic chirality plays a key role there. The materials
of this category are not only of keen scientific interest, but they may also
open a possible new window for new device synthesis and fabrication in spintronics.

\begin{acknowledgments}
We acknowledge helpful discussions with J.~Akimitsu, I.~Fomin, Yu.~A.~Izyumov,
and M.~Sigrist. J.~K. acknowledges Grant-in-Aid for Scientific Research
(A)(No.~18205023) and (C) (No.~19540371) from the Ministry of Education,
Culture, Sports, Science and Technology, Japan.
\end{acknowledgments}

\appendix

\section{Periodic potential and Bloch theorem}

We have Schrodinger equation
\begin{equation}
-\frac{d^{2}u}{d{\bar{x}}^{2}}+V({\bar{x}})u=\varepsilon u, \label{Schr}%
\end{equation}
where the periodic potential has a period $2K$
\begin{equation}
V({\bar{x}}+2K)=V({\bar{x}}), \label{prd}%
\end{equation}
and given explicitly by
\begin{equation}
V({\bar{x}})=6\kappa^{2}\mathrm{sn\,}^{2}({\bar{x}})-\kappa^{2}-4+4{\bar
{q}_{0}}\mathrm{dn\,}{\bar{x}}. \label{potent}%
\end{equation}
According to Bloch theorem a class of bounded states is given by
\[
u({\bar{x}})=e^{iQ{\bar{x}}}\phi_{Q}({\bar{x}}),
\]
where $\phi_{Q}({\bar{x}})$ is a periodic function $\phi_{Q}({\bar{x}%
}+2K)=\phi_{Q}({\bar{x}})$ and $Q$ is a Floquet index. It may be shown (see
Ref.{\cite{Flugge}}, for example) that boundary points of bands are determined
from
\[
\cos(2KQ)=\pm1,
\]
that produces boundary points of Brillouin zones
\[
Q_{BZ}^{(n)}=\frac{\pi}{2K}n,\qquad n=\pm1,\pm2,\cdots
\]
The periodicity condition (\ref{prd}) means that the potential may be expanded
into the Fourier series
\[
V({\bar{x}})=\sum_{G_{n}}V_{n}e^{iG_{n}{\bar{x}}},
\]
where the reciprocal lattice points are $G_{n}=2\pi n/(2K)$, $n$ is integer.

To find Fourier coefficients of the potential $V({\bar{x}})$ we use Fourier
series for $\mathrm{dn\,}({\bar{x}})$ and $\mathrm{sn\,}^{2}({\bar{x}})$
functions,
\[
\mathrm{dn\,}({\bar{x}})=\frac{\pi}{2K}+\frac{\pi}{K}\sum_{n=1}^{\infty}%
\frac{\cos{(\pi n{\bar{x}}/K)}}{\cosh{(\pi nK^{\prime}/K)}},
\]
and
\[
\mathrm{sn\,}^{2}({\bar{x}})=\frac{K-E}{K\kappa^{2}}-\sum_{n=1}^{\infty}%
\frac{\pi^{2}n}{\kappa^{2}K^{2}}\frac{\cos{(\pi n{\bar{x}}/K)}}{\sinh{(\pi
nK^{\prime}/K)}}.
\]
Plugging these series into ($\ref{potent}$) we obtain
\[
V_{0}=1+{\kappa^{^{\prime}}}^{2}-6\frac{E}{K}+\frac{2\pi}{K}{\bar{q}_{0}},
\]%
\[
V_{n}=-\frac{3\pi^{2}}{K^{2}}\frac{n}{\sinh{(\pi nK^{\prime}/K)}}+\frac
{2\pi{\bar{q}_{0}}}{K}\frac{1}{\cosh{(\pi nK^{\prime}/K)}}.
\]
The zeroth-order component $V_{0}$ determines a shift and may be omitted while
the component $V_{n}$ mixes the plane waves with wave vectors ${\tilde{k}}$
and ${\tilde{k}}^{\prime}={\tilde{k}}+\pi n/K$
\[
\left\langle {\tilde{k}}^{\prime}|V({\bar{x}})|{\tilde{k}}\right\rangle
=\sum_{n}V_{n}\delta_{{\tilde{k}}^{\prime},{\tilde{k}}+G_{n}}=\sum_{n}%
V_{n}\delta_{{\tilde{k}}^{\prime},{\tilde{k}}+\pi n/K}.
\]
Hence, a quasidegenerate perturbation theory built in the subspace spanned by
two states $|{\tilde{k}}\rangle$ and $|{\tilde{k}}+G_{n}\rangle$%
\begin{equation}
\left.
\begin{array}
[c]{|cc|}%
E_{k}^{0} & V_{n}\\
V_{n}^{\ast} & E_{{\tilde{k}}+G_{n}}^{0}%
\end{array}
\right.  =0,
\end{equation}
yields bands
\[
E_{\pm}({\tilde{k}})=\frac{1}{2}\left(  E_{{\tilde{k}}}^{0}+E_{{\tilde{k}%
}+G_{n}}^{0}\right)  \pm\sqrt{\frac{\left(  E_{{\tilde{k}}}^{0}-E_{{\tilde{k}%
}+G_{n}}^{0}\right)  ^{2}}{4}+|V_{n}|^{2}}.
\]
The gap between the states $|-Q_{BZ}^{(n)}\rangle$ and $|-Q_{BZ}^{(n)}%
+G_{n}\rangle$ is
\[
2|V_{n}|=\left\vert -\frac{6\pi^{2}}{K^{2}}\frac{n}{\sinh{(\pi nK^{\prime}%
/K)}}+\frac{4\pi{\bar{q}_{0}}}{K}\frac{1}{\cosh{(\pi nK^{\prime}/K)}%
}\right\vert ,
\]
and it falls rapidly to zero with increasing of $n$
\[
2|V_{n}|\simeq\exp{(-\pi nK^{\prime}/K)}.
\]

\section{LAM$\acute{\mathrm{E}}$ EQUATION}

The basic properties of the Lam$\acute{\mathrm{e}}$ equation are presented
here. We start with the Jacobi form which is defined by \cite{WW}
\begin{equation}
{\frac{d^{2}{\Lambda}_{{\bar{\alpha}}}(x)}{dx^{2}}}=\left[  \ell(\ell
+1)\kappa^{2}\mathrm{sn\,}^{2}\left(  x,\kappa\right)  -\kappa+^{2}%
(1+A)\right]  {\Lambda}_{{\bar{\alpha}}}(x),
\end{equation}
where $\ell=1$ and $A$ being a constant. The spectrum is labeled by a complex
parameter ${\bar{\alpha}}$ and given by
\begin{equation}
A_{\bar{\alpha}}={\frac{1}{\kappa^{2}}}\mathrm{dn\,}^{2}\bar{\alpha}.
\end{equation}
The solution of the Lam$\acute{\mathrm{e}}$ equation is exactly given in the
conventional form \cite{WW}
\begin{equation}
\Lambda_{{\bar{\alpha}}}(x)={\frac{\mathrm{H}(x-\bar{\alpha})}{\Theta(x)}%
}e^{xZ(\bar{\alpha})}, \label{conventional_rep}%
\end{equation}
where $\mathrm{H}$, $\Theta,$ and $Z$ are Jacobi's eta, theta, and zeta
functions, respectively, with the elliptic modulus $\kappa.$ Now, we require
Eq.(\ref{conventional_rep}) to be a propagating Bloch wave, i.e.,
$Z(\bar{\alpha})$ to be pure imaginary. Recalling that the zeta function
$Z(\bar{\alpha})$ is singly periodic with the period \ $2K,$ we see that two
segments $(K-2iK^{\prime},K]$ and $[-2iK^{\prime},0)$ for $\bar{\alpha}$ are
sufficient to fully describe the solution (\ref{conventional_rep}).

Because of the quasi-periodicity,
\[
H(x+2K-\bar{\alpha})=-H(x-\bar{\alpha}),\,\,\Theta(x+2K)=\Theta(x),
\]
we have
\[
\Lambda_{\bar{\alpha}}(x+2K)=-e^{2KZ(\bar{\alpha})}\Lambda_{\bar{\alpha}}(x),
\]
and it is convenient to introduce the Floquet index
\[
\bar{Q}(\bar{\alpha})={\frac{\pi}{2K}}+iZ(\bar{\alpha},k).
\]
Then, we have
\[
\Lambda_{\bar{\alpha}}(x+2K)=e^{-2Ki\bar{Q}}\Lambda_{\bar{\alpha}}(x),
\]
that is analogous to the Bloch theorem where $2K$ and $\bar{Q}$ have the
meanings of the lattice constant and the quasimomentum, respectively.
Furthermore, imposing the periodic boundary condition
\begin{align}
\Lambda_{\bar{\alpha}}(x+L)  &  =\Lambda_{\bar{\alpha}}\left(  x+{\frac{L}%
{2K}}2K\right)  =\left[  e^{-2Ki\bar{Q}}\right]  ^{\frac{L}{2K}}\Lambda
_{\bar{\alpha}}(x)\nonumber\\
&  =e^{-iL\bar{Q}}\Lambda_{\bar{\alpha}}(x)=\Lambda_{\bar{\alpha}}(x),
\end{align}
we have the quasi-momentum as usual,
\[
\bar{Q}={\frac{2\pi}{L}}n,
\]
where $n$ is integer.

Finally, we have the Bloch form,
\[
\Lambda_{\bar{\alpha}}(x)={\frac{\mathrm{H}(x-\bar{\alpha})}{\Theta(x)}%
}e^{-i\bar{Q}x}e^{i{\frac{\pi}{2K}}x}.
\]
Other than the conventional parameterization, it is convenient to work with a
real parameter ${\alpha}$ related with $\bar{\alpha}$ by
\begin{equation}
\bar{\alpha}=i\alpha+K-iK^{\prime}, \label{transf}%
\end{equation}
for the acoustic branch, and
\begin{equation}
\bar{\alpha}=i\alpha-iK^{\prime},
\end{equation}
for the optic one. Within the new parametrization the eigenfunction for the
acoustic mode transforms in the following way
\begin{align*}
&  H(x-i\alpha-K+iK^{\prime})\\
&  =\vartheta_{1}\left(  {\frac{\pi}{2K}}[x-x_{0}+iK^{\prime}]\right) \\
&  =ie^{\frac{\pi K^{\prime}}{4K}}e^{i{\frac{\pi}{2K}}x_{0}}e^{-i{\frac{\pi
}{2K}}x}\vartheta_{4}\left(  {\frac{\pi}{2K}}[x-x_{0}]\right)  ,
\end{align*}
where $x_{0}=i\alpha+K$, and $\vartheta_{i}$ ($i=1,2,3,4$) denote the Theta
functions. Furthermore, we have
\[
\vartheta_{3}\left(  {\frac{\pi}{2K}}[x_{0}-iK^{\prime}]\right)  =e^{\frac{\pi
K^{\prime}}{4K}}e^{i{\frac{\pi}{2K}}x_{0}}\vartheta_{2}\left(  {\frac{\pi}%
{2K}}x_{0}\right) ,
\]
and
\[
{\mathrm{H}(x-\bar{\alpha})}=i{\frac{\vartheta_{3}\left(  {\dfrac{\pi}{2K}%
}[x_{0}-iK^{\prime}]\right)  }{\vartheta_{2}\left(  {\dfrac{\pi}{2K}}%
x_{0}\right)  }}e^{-i{\frac{\pi}{2K}}x}\vartheta_{4}\left(  {\frac{\pi}{2K}%
}[x-x_{0}]\right)  ,
\]
that yields
\begin{align}
\label{Izyumovrep}{\Lambda}_{{\alpha}}(x) =i{\dfrac{\vartheta_{3}\left(
{\dfrac{\pi}{2K}}[x_{0}-iK^{\prime}]\right)  }{\vartheta_{2}\left(
{\dfrac{\pi}{2K}}x_{0}\right)  }}{\dfrac{ \vartheta_{4}\left(  {\dfrac{\pi
}{2K}}[x-x_{0}]\right)  }{\vartheta_{4}\left(  {\dfrac{\pi}{2K}}x\right)  }%
}e^{-i\bar{Q}x}.\nonumber\\
\end{align}
This is an alternative representation for the solution (\ref{conventional_rep}%
),\cite{Izyumov-Laptev86} and it is used in the paper. The case of the optic
branch ($x_{0}=i\alpha$) is considered by a similar way.

The transformation of the Floquet index for the acoustic branch is carried out
as follows. By noticing that
\begin{align*}
&  Z(i\alpha+K-iK^{\prime})\\
&  =Z(i\alpha)+Z(K-iK^{\prime})-\mathrm{sn\,}(i\alpha)\mathrm{dc}(i\alpha)\\
&  =Z(i\alpha)+i{\frac{KE^{\prime}+K^{\prime}E-KK^{\prime}}{K}}%
-\mathrm{\ sn\,}(i\alpha)\mathrm{dc}(i\alpha)\\
&  =i{\frac{\pi}{2K}}-iZ(\alpha,k^{\prime})-i\pi{\frac{\alpha}{ 2KK^{\prime}}%
},
\end{align*}
where we used the Jacobi's imaginary transformations and the Legendre's
relation $KE^{\prime}+K^{\prime}E-KK^{\prime}=\pi/2$. Therefore, we have
\[
\bar{Q}(\bar{\alpha})=Q(\alpha)={\dfrac{\pi\alpha}{2KK^{\prime}}}+Z(
\alpha,\kappa^{\prime}).
\]
The same transformation for the optic mode (${\bar\alpha}=i \alpha- i
K^{\prime}$) yields
\begin{equation}
Q(\alpha)= {\dfrac{\pi\alpha}{2KK^{\prime}}}+Z\left(  \alpha,\kappa^{\prime
}\right)  +\mathrm{dn\,}\left(  \alpha,\kappa^{\prime}\right)  {\dfrac
{\mathrm{cn\, }\left(  \alpha,\kappa^{\prime}\right)  }{\mathrm{sn\,}\left(
\alpha,\kappa^{\prime}\right)  }.}%
\end{equation}

By the same manner, the corresponding spectrum is parametrized as
\[
\bar{A}_{\bar{\alpha}}={\dfrac{1}{\kappa^{2}}}\mathrm{dn\,} ^{2}{\bar\alpha}
={A}_{{\alpha}}= \left\{
\begin{array}
[c]{c}%
{\dfrac{\kappa^{\prime2}}{\kappa^{2}}}\mathrm{sn\,}^{2}{\alpha} \,\,\,
({\text{acoustic}})\\
{\dfrac{1}{\kappa^{2}\mathrm{sn\,}^{2}{\alpha}}} \,\,\, ({\text{optic}})
\end{array}
. \right.
\]

Now, we briefly review the origin of the band structure.\cite{Sutherland73} In
the limit $\kappa\rightarrow1,$ the Lam$\acute{\mathrm{e}}$ equation reduces
to the Schr\"{o}dinger equation,
\[
{\frac{d^{2}\varphi(x)}{dx^{2}}}+E+U_{0}\mathrm{\,sech\,}^{2}(\alpha x)=0,
\]
where $E=k^{2}(1+A)-\ell(\ell+1),\,\,\,\,U_{0}=\ell(\ell+1).$The potential
\[
U(x)=-U_{0}\mathrm{\,sech\,}^{2}(\alpha x),
\]
is modified P\"{o}schl-Teller potential and for $\ell=1,$ there are one bound
state and one \textit{perfectly transmitted (reflectionless) }{\ \ }scattering
state.\cite{LL} The band structure of the Lam$\acute{\mathrm{e}}$ equation is
understood as follows. In the limit of well separated modified
P\"{o}schl-Teller potential, the $\ell$ bound states give discrete levels and
the scattering states give broad continuum. When the potentials form a
lattice, the discrete level overlaps and the energy band may be formed. Even
after the band formation, the gap between the bound level and the scattering
continuum retains. Therefore, the resulting band is split into the lower
acoustic band and the upper optical band.

\section{Dirac's canonical formulation for the singular Lagrangian theory}

The canonical momenta conjugate to the coordinates $\overline{X}(t),$
$\eta_{\alpha}(t),$ and $\xi_{\alpha}(t)$ are given by
\begin{align}
\left\{
\begin{array}
[c]{c}%
p_{1}={\partial}{{\mathcal{L}}}/{\partial}\dot{q}_{1}=c_{0}\sum_{\alpha
}{\mathcal{K}}_{\alpha}q_{3\alpha},\\
p_{2\alpha}={\partial}{{\mathcal{L}}}/{\partial}\dot{q}_{2\alpha}%
=-c_{0}\left(  {\mathcal{J}}_{\alpha}+\sum_{\beta}{\mathcal{M}}_{\alpha\beta
}q_{3\beta}\right)  ,\\
p_{3\alpha}={\partial}{{\mathcal{L}}}/{\partial}\dot{q}_{3\alpha}=0,
\end{array}
\right.
\end{align}
and we obtain a canonical Hamiltonian,
\begin{equation}
H_{\text{c}}=p_{1}\dot{q}_{1}+\sum_{\alpha}p_{2\alpha}\dot{q}_{2\alpha}%
+\sum_{\alpha}p_{3\alpha}\dot{q}_{\alpha}-{{\mathcal{L}}}.
\end{equation}
The Lagrangian (\ref{Lagrangian}) itself gives rise to a set of primary
constraints,
\begin{align}
\left\{
\begin{array}
[c]{l}%
\phi_{1}^{(1)}=p_{1}-c_{0}\sum_{\alpha}{\mathcal{K}}_{\alpha}q_{3\alpha
}\approx0,\\
\phi_{2\alpha}^{(1)}=p_{2\alpha}+c_{0}\left(  {\mathcal{J}}_{n}+\sum_{\beta
}{\mathcal{M}}_{\alpha\beta}q_{3\beta}\right)  \approx0,\\
\phi_{3\alpha}^{(1)}=p_{3\alpha}\approx0,
\end{array}
\right.
\end{align}
where the symbol $\approx0$ means "weakly zero," i.e. $\phi_{i}^{(1)}$ may
have nonvanishing canonical Poisson brackets with some canonical variables.
Because of a lack of primary expressible velocities the Hamiltonian with the
imposed constraints,%
\begin{equation}
H^{\ast}=\phi_{1}^{(1)}\dot{q}_{1}+\sum_{\alpha}\phi_{2\alpha}^{(1)}\dot
{q}_{2\alpha}+\sum_{\alpha}\phi_{3\alpha}^{(1)}\dot{q}_{3\alpha}%
+c_{1}{\mathcal{V}},
\end{equation}
coincides with $H_{\text{c}}$, i.e. $\dot{q}_{1},$ $\dot{q}_{2\alpha},$ and
$\dot{q}_{3\alpha}$ (primary inexpressible velocities) play the role of
Lagrangian multipliers. Now, the Hamiltonian $H^{\ast}$ governs the equations
of motion of the constrained system. The relevant non-zero Poisson brackets
are computed as,%
\begin{align}
\left.
\begin{array}
[c]{c}%
\left\{  \phi_{1}^{(1)},\phi_{3\alpha}^{(1)}\right\}  =-c_{0}{\mathcal{K}%
}_{\alpha},\\
\left\{  \phi_{2\alpha}^{(1)},\phi_{3\alpha}^{(1)}\right\}  =c_{0}\sum_{\beta
}{\mathcal{M}}_{\alpha\beta},\\
\left\{  \phi_{2\alpha}^{(1)},{\mathcal{V}}\right\}  =-2c_{1}\rho_{\alpha
}q_{2\alpha},\\
\left\{  \phi_{3\alpha}^{(1)},{\mathcal{V}}\right\}  =-2c_{1}\lambda_{\alpha
}q_{3\alpha},
\end{array}
\right\}
\end{align}
and $\left\{  q_{i},p_{j}\right\}  =\delta_{ij}$ gives rise to the constraint
conditions, $\dot{\phi}_{1}^{(1)}=\left\{  {\phi}_{1}^{(1)},H^{\ast}\right\}
=0$, $\dot{\phi}_{2\alpha}^{(1)}=\left\{  {\phi}_{2\alpha}^{(1)},H^{\ast
}\right\}  =0$, and $\dot{\phi}_{3\alpha}^{(1)}=\left\{  {\phi}_{3\alpha
}^{(1)},H^{\ast}\right\}  =0$, or in the explicit form%

\begin{align}
\dot{\phi}_{1}^{(1)}  &  =c_{0}\sum_{\alpha}{\mathcal{K}}_{\alpha}\dot
{q}_{3\alpha}=0,\;\;\;\;\;\;\label{c1}\\
\dot{\phi}_{2\alpha}^{(1)}  &  =c_{0}\sum_{\beta}{\mathcal{M}}_{\alpha\beta
}\dot{q}_{3\beta}-2c_{1}\rho_{\alpha}q_{2\alpha}=0,\;\;\;\;\;\;\label{c2}\\
\dot{\phi}_{3\alpha}^{(1)}  &  =c_{0}\left(  {\mathcal{K}}_{\alpha}\dot{q}%
_{1}-\sum_{\beta}{\mathcal{M}}_{\alpha\beta}\dot{q}_{2\alpha}\right)
-2c_{1}\lambda_{\alpha}q_{3\alpha}\nonumber\\
&  =0.\qquad\label{c3}%
\end{align}
Eq. (\ref{c1}) gives $\dot{q}_{3\alpha}=0$ and then Eq. (\ref{c2}) gives
$q_{2\alpha}=0$. Now, there arises the secondary constraints $\phi_{\alpha
}^{(2)}=q_{2\alpha}\approx0$ to be constant in time, $\dot{\phi}_{\alpha
}^{(2)}$ $=\left\{  {\phi}_{\alpha}^{(2)},H^{\ast}\right\}  $ $=\dot
{q}_{2\alpha}=0$, and the consistency condition is fulfilled. Finally Eq.
(\ref{c3}) relates $q_{3\alpha}=\xi_{\alpha}$\ to $\dot{q}_{1}=\dot{X}$, and
produces Eq.(\ref{canonicalrelation1}).

\section{Computation of ${\mathcal{K}}_{\alpha}$}

We compute
\begin{align*}
{\mathcal{K}}_{\alpha}  &  =2\int_{0}^{L}dx\mathrm{\,dn\,}\left(
x,\kappa\right)  u_{\alpha}\left(  x\right) \\
&  =2N(\alpha)\int_{0}^{L}dx\mathrm{\,dn\,}\left(  x,\kappa\right)
{\frac{\vartheta_{4}\left(  \dfrac{\pi}{2K}(x-x_{0})\right)  }{\vartheta
_{4}\left(  \dfrac{\pi}{2K}x\right)  }}e^{-i{Q}x},
\end{align*}
with $x_{0}=i\alpha+K$, and where $N\left(  \alpha\right) $ is a normalization
factor. Noting that $\mathrm{dn\,}\left(  x,\kappa\right)  {\frac
{\vartheta_{4}\left(  \frac{\pi}{2K}(x-x_{0})\right)  }{\vartheta_{4}\left(
\frac{\pi}{2K}x\right)  }}$ has a period 2$K$, we perform the Fourier
decomposition,
\[
\mathrm{dn\,}\left(  x,\kappa\right)  {\frac{\vartheta_{4}\left(  \frac{\pi
}{2K}(x-x_{0})\right)  }{\vartheta_{4}\left(  \frac{\pi}{2K}x\right)  }=}%
\sum_{l}\gamma_{l}e^{i\frac{\pi x}{K}l},
\]
where the coefficients are evaluated as
\[
\gamma_{l}=\frac{1}{2K}\int_{-K}^{K}dx\mathrm{dn\,}\left(  x,\kappa\right)
{\frac{\vartheta_{4}\left(  \frac{\pi}{2K}(x-x_{0})\right)  }{\vartheta
_{4}\left(  \frac{\pi}{2K}\tilde{x}\right)  }}e^{-^{i\frac{\pi x}{K}l}}.
\]
Then, we have
\begin{align*}
{\mathcal{K}}_{\alpha}  &  = 2{N}\left(  \alpha\right)  L\sum_{l}\gamma
_{l}\delta_{Q,\frac{\pi}{K}l}.
\end{align*}
Within the acoustic branch ($0\leq|Q|\leq{\frac{\pi}{2K}}$), only $Q=0$
($\alpha=0$) contributes to ${\mathcal{K}}_{\alpha}$. Eventually, the
orthogonality condition (\ref{orthogonalilty})\ of a denumerable basis
enforces that there is no contribution of the term with $l\neq0$. By using
${\Lambda}_{{\alpha=0}}(x)=\sqrt{\frac{K\left(  \kappa\right)  }{LE\left(
\kappa\right)  }}\mathrm{dn\,}\left(  x,\kappa\right) $, therefore we have
${\mathcal{K}}_{\alpha}=\delta_{\alpha,0}{\mathcal{K}}_{0}$,where%
\begin{align*}
{\mathcal{K}}_{0}  &  =2\sqrt{\frac{K\left(  \kappa\right)  }{E\left(
\kappa\right)  \bar{L}}}\int_{0}^{\bar{L}}\mathrm{\,dn}^{2}\mathrm{\,}\left(
x,\kappa\right)  \mathrm{\,}dx =2\sqrt{\frac{E\left(  \kappa\right)
}{K\left(  \kappa\right)  }\bar{L}},
\end{align*}
where we exploited the relation $E\left(  \kappa\right)  =\int_{0}^{K\left(
\kappa\right)  }\mathrm{\,dn}^{2}\mathrm{\,}\left(  x,\kappa\right)
\mathrm{\,}dx$.

\section{Inertial motion of Bloch wall}

We here discuss the relevance of the present formulation to the
D\"{o}ring-Becker-Kittel mechanism.\cite{Doring48,Becker50,Kittel50} We
consider a conventional Bloch wall in ferromagnets, where the magnetization
rotates through the plane of the wall. The wall size is determined by the
exchange energy cost and the anisotropy energy that amount to%
\begin{equation}
\sigma=\frac{\pi^{2}JS^{2}}{Na_{0}^{2}}+KNa_{0},
\end{equation}
where $N$\ is the number of spins inside the wall. Minimizing this energy
leads to the wall size $l_{\text{Bloch}}=\pi\sqrt{JS^{2}/Ka_{0}}$ with
$K$\ denoting the anisotropy energy. Now, let us consider the Bloch wall
formed along the $x$-axis and spins are confined to the $yz$-plane that winds
180$^{\circ}$. D\"{o}ring proposed that the translation of the domain wall is
driven by the appearance of the local demagnetization field $H_{x}$ inside the
wall that violates the condition $\mathbf{\nabla\cdot}{\mathbf{M}=0}$, i.e.
$H_{x}=-4\pi\left[  M_{x}-M_{x}(\infty)\right]  $, and causes the precessional
motion of the magnetization within the $yz$-plane. Then, the corresponding
Larmor frequency amounts to $\omega_{\text{L}}=\dot{\varphi}=\gamma H_{x}$,
where $\gamma$ is a gyromagnetic ratio. On the other hand, in the steady
movement of the wall, $\dot{\varphi}=-\left(  \partial_{x}\varphi\right)  V$,
with $V$ being the velocity, and consequently we have
\begin{equation}
H_{x}=-\gamma^{-1}\left(  \partial_{x}\varphi\right)  V. \label{Hvel}%
\end{equation}
The excess of magnetization energy
\[
\Delta W=\frac{1}{8\pi}\int_{-\infty}^{\infty}H_{x}^{2}dx=\frac{V^{2}}%
{8\pi{\gamma}^{2}}\int_{-\infty}^{\infty}\left(  \frac{\partial\varphi
}{\partial x}\right)  ^{2}dx
\]
gives the energy stored in the moving wall. Taking the form $\Delta
W=M_{\text{D\"{o}ring}}V^{2}/2$ the inertial mass of the wall first proposed
by D\"{o}ring\cite{Doring48} is introduced
\begin{equation}
M_{\text{D\"{o}ring}}=\frac{1}{4\pi{\gamma}^{2}}\int_{-\infty}^{\infty}\left(
\frac{\partial\varphi}{\partial x}\right)  ^{2}dx. \label{DoringMass}%
\end{equation}
The explicit form $\left(  \partial\varphi/\partial x\right)  ^{2}$ depends on
kind of domain walls, their orientation around crystallographic axes, for
example
\[
M_{\text{D\"{o}ring}}=\frac{1}{4\pi{\gamma}^{2}}\sqrt{K/J}%
\]
for $180$-degree domain wall parallel to crystallographic plane (100). Taking
into account that $\gamma=1.84\times10^{7}\,$ $[(\text{Oe}\cdot\text{s}%
)^{-1}]$ this yields in the case of Fe
\[
M_{\text{D\"{o}ring}}\simeq10^{-10}[\text{g}/\text{cm}^{2}].
\]

>From Eq. (\ref{Hvel}) it stems that
\[
M_{x}=-\frac{1}{4\pi}H_{x}=\frac{1}{4\pi\gamma}\left(  \partial_{x}%
\varphi\right)  V
\]
provided $M_{x}(\infty)=0$. This equation should be compared with
Eq.(\ref{NetMag}) rewritten in the form
\[
m(\bar{x})\simeq-\frac{g\mu_{B}\hbar}{2Ja_{0}q_{0}}\,\left(  \partial_{\bar
{x}}\varphi\right)  \,\dot{X}.
\]
We see that $m(\bar{x})$ may be interpreted as the demagnetization field and
the physical Hamiltonian (\ref{physicalHamiltonian}) may be regarded as the
energy cost associated with the demagnetization process.

\end{document}